\documentclass[aps,prb,twocolumn,superscriptaddress,10pt]{revtex4-2}
\usepackage{graphicx}
\usepackage{braket}
\usepackage[breaklinks,unicode=true,colorlinks=true,citecolor=blue,urlcolor=blue]{hyperref} 
\usepackage{amsmath}
\usepackage[dvipsnames]{xcolor} 

\usepackage{bm}

\begin{document}

\newcommand{\thetitle}{Magnetic switching of exciton lifetime in CrSBr}
\title{\thetitle}

\author{Ina V. Kalitukha*} 
\author{Ilya A. Akimov}
\author{Mikhail O. Nestoklon}
\affiliation{Experimentelle Physik 2, Technische Universit\"{a}t Dortmund, 44227 Dortmund, Germany}
\author{Torsten Geirsson}
\affiliation{Institute of Materials Science (ICMUV), University of Valencia, Catedrático Beltrán 2, E-46980 Valencia, Spain}
\author{Alejandro Molina-S\'anchez}
\affiliation{Institute of Materials Science (ICMUV), University of Valencia, Catedrático Beltrán 2, E-46980 Valencia, Spain}
\author{Ey\"up Yalcin}
\author{Claudia Ruppert}
\affiliation{Experimentelle Physik 2, Technische Universit\"{a}t Dortmund, 44227 Dortmund, Germany}
\author{Daniel A. Mayoh}
\author{Geetha Balakrishnan}
\affiliation{Department of Physics, University of Warwick, Coventry CV4 7AL, United Kingdom}
\author{Muthumalai Karuppasamy}
\author{Zden\v{e}k Sofer}
\affiliation{Department of Inorganic Chemistry, University of Chemistry and Technology Prague, Technicka 5, 166 28 Prague 6, Czech Republic}
\author{Yadong Wang}
\author{Daniel J. Gillard}
\author{Xuerong Hu}
\author{Alexander I. Tartakovskii}
\affiliation{Department of Physics and Astronomy, The University of Sheffield, Sheffield S3 7RH, United Kingdom}
\author{Manfred Bayer}
\affiliation{Experimentelle Physik 2, Technische Universit\"{a}t Dortmund, 44227 Dortmund, Germany}
\affiliation{Research Center FEMS, Technische Universit\"{a}t Dortmund, 44227 Dortmund, Germany}

\begin{abstract} 
Exciton dynamics in layered magnetic semiconductors provide a sensitive probe of the interplay between spin order and light--matter interaction. Here, we study thin CrSBr layers using time-resolved photoluminescence spectroscopy in an external magnetic field, revealing a step-like reduction in the exciton lifetime from 11 to 7~ps, during the magnetization flip from the antiferromagnetic to the ferromagnetic phase. The reduction of the exciton lifetime in the ferromagnetic phase persists below the N\'eel temperature, as evidenced by its strong magnetic-field dependence that disappears in the paramagnetic phase. {\it Ab initio} calculations reveal a one-dimensional nature of free excitons accompanied by a pronounced change in the oscillator strength across the magnetic phase transition predicting a shorter radiative lifetime of free excitons in the antiferromagnetic phase of CrSBr contradicting the experimental observations. This discrepancy is explained by strong localization of excitons at low tempature. We show both experimentally and theoretically that the observed magnetic switching of the exciton lifetime is attributed to a larger exciton localization volume leading to a larger oscillator strength in the ferromagnetic phase.
The results show that disorder-induced localization effects play a key role in exciton dynamics in CrSBr.
\end{abstract}

\maketitle

\section{Introduction} 

Two-dimensional (2D) van der Waals (vdW) materials possess unique electronic band structure and  optical properties, being attractive for a wide variety of optoelectronic, sensing and quantum technology applications~\cite{Novoselov-review2016}. Magnetic vdW materials attract special attention for the development of spintronics devices and for investigation of magnetic interactions in strongly correlated low-dimensional systems~\cite{Wang-review2022}. CrSBr is a special case of an air-stable, two-dimensional magnetic semiconductor that can be tuned in a moderate magnetic field between the ferromagnetic (FM) and antiferromagnetic (AFM) phases at temperatures below 132~K \cite{CrSBr-review2024, Maletinski-2024}. The magneto-optical response is dominated by the Wannier-Mott exciton exhibiting giant exchange interaction with magnetic Cr ions~\cite{Wilson-2021, Haegele-2024}, making  CrSBr a unique magnetic semiconductor in which excitons can be used as a probe of coherent magnon dynamics~\cite{Zhu-magnons-2022, Dirnberger-2023}. Furthermore, efficient light-matter coupling with prominent exciton-polariton features was reported in CrSBr flakes embedded in planar microcavity~\cite{Wang2023}. Such a combination of optical and magnetic properties is appealing for realization of optical control of magnetic order via excitons.

Excitons provide optical access to the electronic structure in semiconductor nanostructures~\cite{Ivchenko-book}. In particular, the exciton oscillator strength and population dynamics depend sensitively on the material's dimensionality, disorder, and purity. Several studies pointed out that in bulk CrSBr, the dimensionality of conduction band electrons changes from quasi-one-dimensional (1D) to two-dimensional (2D) behavior during the transition from AFM to FM phase~\cite{Klein-2023, Telford-2022}. Such a transition should modify the exciton wavefunction and its radiative lifetime. However, the intrinsic exciton dynamics in CrSBr and its dependence on the magnetic phase are relatively unexplored. Recently, THz spectroscopy revealed ultrafast exciton dynamics in CrSBr down to the monolayer limit~\cite{Meineke-2024}. The authors evaluated the lifetime of 15~ps in the bulk and a significantly shorter time of 0.5~ps in the monolayer. Time-resolved photoluminescence (TRPL) studies reported a similar value of about 20~ps in bulk CrSBr layers~\cite{Lin-2024}. Yet non-radiative channels could not be excluded as a source of exciton lifetime shortening in a CrSBr monolayer. Moreover, previous studies of exciton dynamics were performed at zero magnetic field and did not address the radiative lifetime of excitons in different magnetic phases or the modification of the exciton wavefunction in CrSBr.

In this work, we report on exciton dynamics in thin CrSBr layers with thicknesses around 30 to 70~nm using time-resolved photoluminescence spectroscopy in  external magnetic fields $B$ up to $3$~T. At a low temperature of $T=2$~K, we measure the decay of the PL signal and evaluate the exciton lifetime in the order of 10~ps. We demonstrate a step-like reduction in exciton lifetime, from 11 to 7~ps during the magnetization flip from AFM to FM phase for $\mathbf{B}$ directed along the easy magnetic axis.  For other orientations of magnetic field we observe gradual shortening that correlates with the magnetization of the layer. Numerical solutions of the Bethe-Salpeter equation based on density functional theory (DFT) calculations predict the opposite behavior for the radiative lifetime of free excitons in different magnetic phases of CrSBr. We observe an abrupt slowdown of exciton dynamics in both the AFM and FM phases at elevated temperatures, with lifetimes reaching up to 50~ps near the N\'{e}el temperature of $T_{\rm N} = 132$~K. This behavior excludes the non-radiative decay channel at low temperatures and suggests the activation of free excitons that populate wave-vector components outside the light cone. We therefore conclude that at low temperatures below 100~K the PL decay is mainly determined by the radiative emission of localized one-dimensional excitons. In this case, the exciton localization volume, and consequently the oscillator strength, increase during the transition from the AFM to the FM phase due to the reduction of confinement within the van der Waals layers. From the temperature dependence of the lifetime in the AFM and FM phases we estimate the change of localization energy from 40 to 32~meV, respectively. Our results demonstrate the crucial role of disorder-induced localization effects on exciton dynamics in CrSBr.

\section{Experimental results}
\subsection{Time-integrated PL}

\begin{figure}[h!]
	\centering
	\includegraphics[width=1\linewidth]{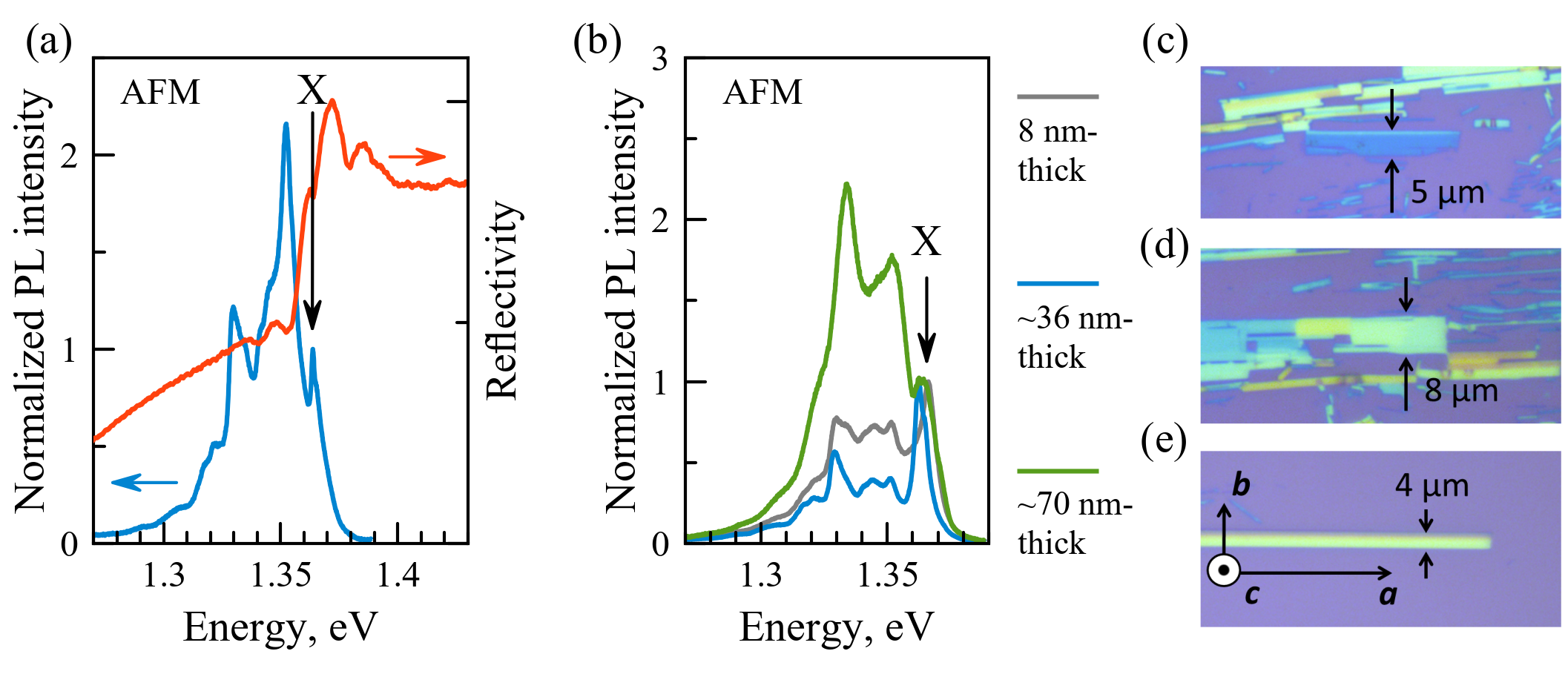}
	\caption{(a) PL and reflectivity spectra in the AFM phase of a characteristic flake in zero magnetic field.   Temperature $T = 2$~K. PL is excited with photon energy $E_{\rm exc} = 1.771$~eV. PL spectrum is normalized to the exciton peak (X). (b) PL spectra in the AFM phase from flakes of different thicknesses (as shown in c-e).  The arrow marks the spectral position of 1s exciton. $T = 2$~K, $E_{\rm exc} = 1.771$~eV. PL intensity is normalized by the X peak intensity. (c), (d) and (e) Optical microscope images of flakes with 8, $\sim$36 and $\sim$70~nm   thicknesses, correspondingly, which spectra are shown in (b). Thicknesses are measured using atomic force microscopy, see Supplementary Information Fig.~S1. The thicknesses $\sim$36 and $\sim$70~nm are derived as mean values across the flakes.    Directions of CrSBr $a$, $b$ and $c$ axes are shown by arrows in (e), the same directions are applicable to (c) and (d).}
	\label{fig0}
\end{figure}

Optical properties of different CrSBr flakes in the AFM phase at $B=0$ are summarized in Fig.~\ref{fig0}. Both PL and reflectivity spectra show several features at photon energies around 1.35~eV [see Fig.~\ref{fig0}(a)]. All of them are observed in linear polarization along the CrSBr crystallographic $b$-axis. Therefore, the corresponding linear polarization was used for excitation and detection. The main feature in the reflectivity spectrum is associated with a step-like change of intensity, which occurs around 1.37~eV [see vertical arrow in Fig.~\ref{fig0}(a)]. Its position is close to the high energy peak in the PL spectrum. There is a multitude of PL peaks at lower energies whose relative intensity varies strongly from one flake to another. This is demonstrated in Fig.~\ref{fig0}(b), which presents PL spectra from flakes of different thicknesses ranging from 8 to $\sim$70~nm. The optical microscope images of the corresponding flakes are shown in Figs.~\ref{fig0}(c-e). 

The multitude of PL lines in bulk CrSBr layers of different thicknesses has been discussed in connection with energy band splittings~\cite{Lin-2024b} and exciton-polariton states~\cite{Dirnberger-2023}. We note, however, that although the shape of the PL spectrum varies from flake to flake, all the peaks follow the same energy shift in magnetic field, and their temporal dynamics are nearly identical. Therefore, we attribute the high energy peak in PL at 1.37~eV to the ground state of 1s exciton, and the lower energy peaks to phonon replica, as discussed recently in Refs.~\onlinecite{Lin-2024,Urbaszek-2025}. Such scenario is  also established in polar II-VI semiconductors with strong exciton-phonon interactions~\cite{Permogorov-Excitons}.  

\begin{figure}[h!]
	\centering
	\includegraphics[width=1\linewidth]{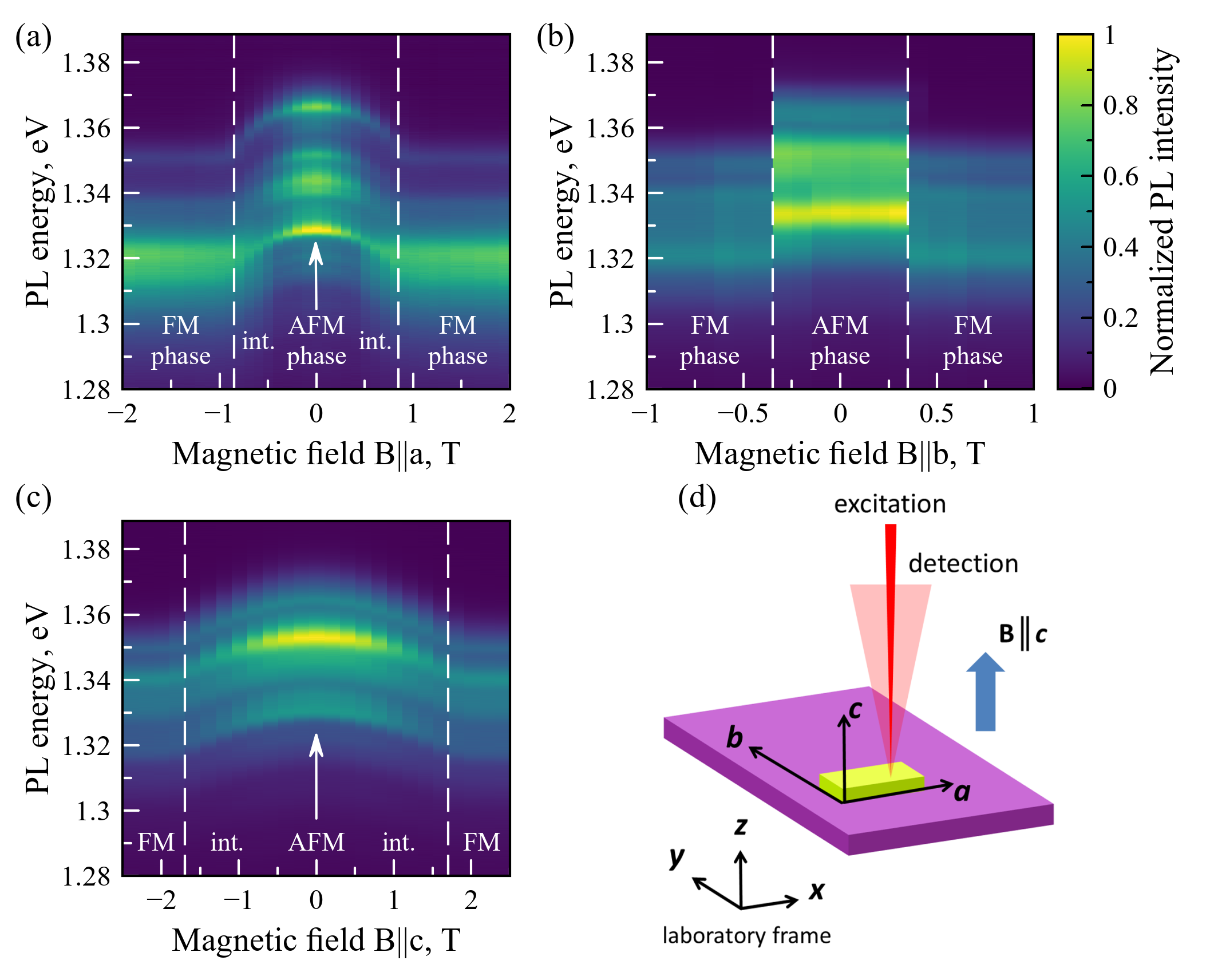}
	\caption{2D plots of PL signal as a function of magnetic field and photon energy measured at $T = 2$~K on a characteristic flake (thickness of $\sim$70~nm), $E_{\rm exc} = 1.771$~eV. (a), (b) Voigt geometry with $\textbf{B} \parallel a$ and $\textbf{B} \parallel b$, respectively. (c) Faraday geometry with $\textbf{B} \parallel c$. (d) Schematic representation of the Faraday experimental geometry. 
	}
	\label{fig1}
\end{figure}

The magnetic field dependence of steady-state PL spectra in different geometries for one of the flakes with the thickness of $\sim$70~nm is shown in Fig.~\ref{fig1}. Slight deviations in PL spectra at $B = 0$ across different panels (a)-(c) are due to inhomogeneities within the flake, such as thickness fluctuations (see Fig.~S1(a) in Supplementary Information SI1), since the measurements are taken at slightly different spots when the geometry was changed. PL spectra reflect the transition from AFM to FM phase under an external magnetic field, which is manifested by reduction of the photon energy of exciton emission by 16~meV. For $\textbf{B}\parallel b$, i.e. when external field is applied along the easy magnetic  axis, there is a sharp transition already at $B=0.35$~T [Fig.~\ref{fig1}(b)], in agreement with previous studies~\cite{Telford-2020, Wilson-2021,Klein-2023s,Haegele-2024}. For magnetic field applied along $a$- and $c$-axis we observe spin canting from the AFM state into the FM state along the direction of magnetic field with saturation fields of 0.85 and 1.7~T, respectively. Similar values were reported for both bulk and few monolayer structures~\cite{Telford-2020, Wilson-2021,Dirnberger-2023,Klein-2023s}. 

One can see in Fig.~\ref{fig1} that the integrated PL intensity is reduced during the transition from AFM into the FM phase, in contrast to PL studies of a few-layer structures in Ref.~\onlinecite{Wilson-2021}. However, it is in line with a previous report on magneto-PL of thin quasi-bulk layers CrSBr in Ref.~\onlinecite{Dirnberger-2023}. Previous studies claimed that intensity changes can be related to the modification of the exciton wavefunction during the phase transition from AFM to FM phase~\cite{Wilson-2021, Klein-2023}. However, the non-radiative processes can also be responsible for the intensity changes. Moreover, the higher energy states of excitons exhibit giant energy shifts in the range of 1.7 to 1.9~eV when a magnetic field is applied, as evidenced by reflectivity spectra measured in Faraday geometry (see Fig.~S2(c) in Supplementary Information SI2). The latter can lead to a variation of absorption as a function of the magnetic field and consequently to a decrease or increase of the total PL signal if the excitation laser is tuned around 1.77~eV, as in our case. The sensitivity of the exciton wavefunction to magnetic field can be revealed in time-resolved PL studies, as shown below.

\subsection{Time-resolved PL}

\begin{figure}[h!]
	\centering
	\includegraphics[width=1\linewidth]{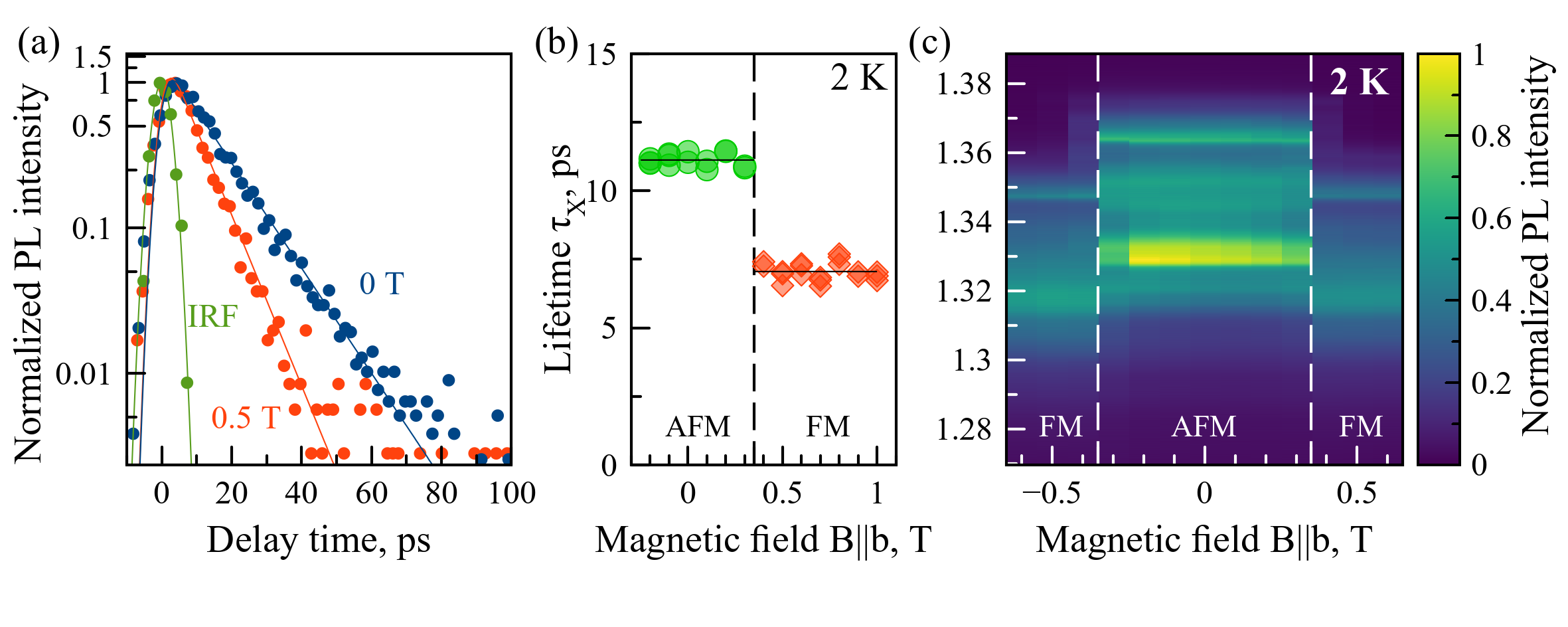}
	\caption{(a) TRPL transients in the AFM ($B =0$, blue) and FM ($B =0.5$~T, red) phases at low temperature ($T = 2$~K). The instrument response function (IRF) is shown by green. Symbols stand for experimental data and solid lines for fit with $I(t)=A\exp{(-\frac{4 \tau_X t - \sigma^2}{4\tau_X^2})}(1+{\rm erf}(\frac{2 \tau_X t-\sigma^2}{2 \tau_X \sigma}))$, representing rise and decay of the PL intensity, where $\tau_X$ is decay time, $A$ is a measure of amplitude of the signal, $\sigma = \rm{HWHM}/\ln 2 \approx 4.3$~ps is the measure of half-width half-maximum (HWHM) of the apparatus function, and ${\rm erf}(t)$ is the error function. (b) Magnetic field dependence of the exciton lifetime $\tau_{\rm X}$ in the AFM (green circles) and FM (red diamonds) phases. The switch of $\tau_{\rm X}$ from $11.1\pm0.3$ to $7.1\pm0.3$~ps takes place during the transition between the AFM and FM phase at $B=0.35$~T. (c) Magnetic field dependence of PL spectrum. Average flake thickness is $\sim$36~nm.
	}
	\label{fig2}
\end{figure}

Time-resolved PL data at low temperature of $2$~K are summarized in Fig.~\ref{fig2}. The decay of the PL intensity does not depend on the emission photon energy, i.e. all PL lines show virtually the same dynamics. Therefore, the signal is integrated in the spectral range of 1.3 to 1.4~eV. Figure~\ref{fig2}(a) shows the resulting intensity transients for $B=0$ and 0.5~T applied along easy magnetic axis ($\textbf{B}\parallel b$). The transients are fitted using the analytical expression obtained from the convolution of a mono‑exponential decay with a Gaussian instrument response function (IRF)~\cite{IRF-2016}. We observe a fast signal rise defined by the total resolution of about 3~ps, see IRF in Fig.~\ref{fig2}(a). The signal decrease follows single exponential decay, which we attribute to exciton lifetime $\tau_{\rm X}$. We note that both the exciton dynamics and the exciton lifetime are independent of the individual flake. The exciton lifetime exhibits only minor flake-to-flake variations ($\le10\%$) and shows no correlation with thickness in the 30–70 nm range.

The magnetic field dependence of lifetime is shown with green circles (AFM phase) and red diamonds (FM phase) in Fig.~\ref{fig2}(b). 
The lifetime switches abruptly from $11.1\pm0.3$ to $7.1\pm0.3$~ps during the magnetization flip from AFM to FM phase, to compare with 2D plot of PL intensity spectra in Fig.~\ref{fig2}(c). Interestingly, the lifetime is shorter in the FM phase. For the magnetic field directed along the hard magnetic $c$-axis, we observe slow shortening to the same value of 7.1~ps for magnetic fields up to 1.7~T [see Fig.~\ref{fig3}(a)]. At higher magnetic fields, once the magnetization is saturated, the exciton lifetime remains constant. The lifetime dependence correlates with the energy position of the exciton PL peak [see Figs.~\ref{fig2}(b) and \ref{fig3}(a)]. Therefore, we conclude that $\tau_{\rm X}$ follows the magnetization and shortens by almost 40\% during the transition from the AFM to the FM phase. To get a deeper insight into the origin of exciton lifetime variation, we study its temperature dependence.

\subsection{Temperature dependence}

\begin{figure}[h!]
	\centering
	\includegraphics[width=1\linewidth]{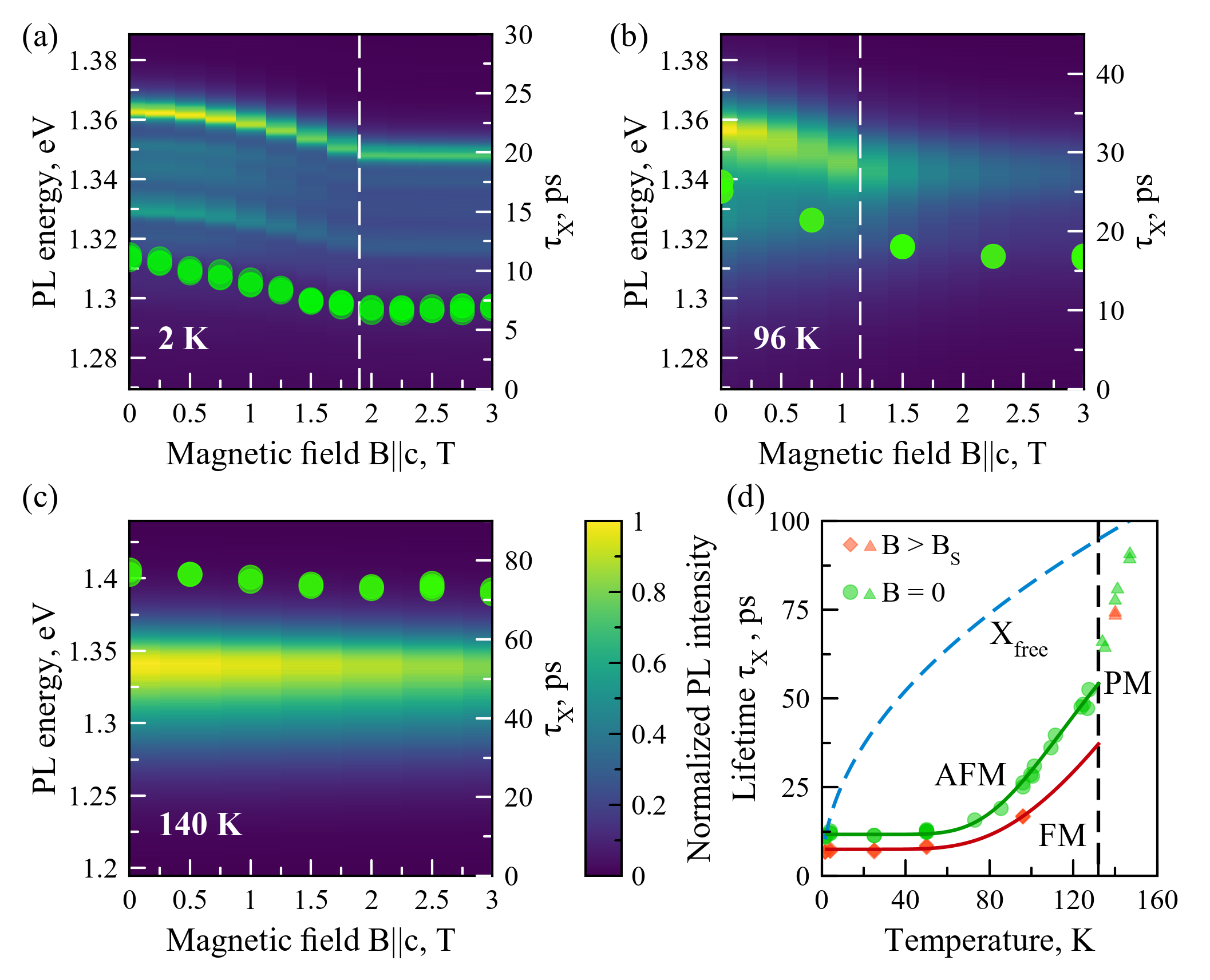}
	\caption{Magnetic field dependence of PL spectra and the exciton lifetime $\tau_{\rm X}$ (green circles) for (a) low temperature $T=2$~K, (b) intermediate temperature $T<T_{\rm N}$ below N\'{e}el temperature, $T=96$~K, and (c) high temperature $T>T_{\rm N}$, $T=140$~K. The data are taken for the magnetic field directed along $c$-axis, in the Faraday geometry. The average flake thickness is $\sim$36~nm. Dashed white line shows transition to the FM phase. (d) Temperature dependence of the exciton lifetime. Green circles show the data at $B=0$ corresponding to the AFM phase. Red diamonds show the data measured in the range of $2-3$~T (larger than the saturation field $B_S$) corresponding to the FM phase. Solid lines are fits with Eq.~\ref{eq:tau_fit} with parameters given in the text. Decay time in the limit of zero temperature  is $\tau_{\rm X}^{\rm AFM}=11.7$~ps in AFM phase and  $\tau_{\rm X}^{\rm FM}=7.5$~ps in FM phase. Black dashed line shows N\'{e}el temperature $T_{\rm N}=132$~K, so that $\tau_{\rm X}$ value above $T_{\rm N}$ (red and green triangles) correspond to paramagnetic phase (PM) disregarding the applied magnetic field. Blue dashed line shows the dependence for the free exciton lifetime $\tau_{\rm FX}$ in 1D case, described in the Discussion section.
    }
	\label{fig3}
\end{figure}

Figures 4(a)-(c) compare the magnetic field dependences of exciton lifetime for $\textbf{B}\parallel c$ (Faraday geometry) measured at temperatures of 2, 96, and 140~K, respectively. For temperatures below the critical N\'{e}el pointOK, corresponding to $T_N=132$~K in bulk CrSBr~\cite{Telford-2020}, we observe similar behavior where $\tau_X$ decreases during the transition from AFM to FM order. For $T=2$~K, $\tau_X$ shows an almost two-fold reduction from 11 to 7~ps (relative change of 36\%) in the range of magnetic fields from 0 to 1.7~T, while for larger $T=96$~K,  $\tau_X$ changes from 25 to 17~ps (relative change of 30\%) for $B \le 1.2$~T. We note that the saturation magnetic field $B_{\rm S}$ required to achieve FM-ordered phase decreases with the temperature increase. In the paramagnetic (PM) phase at $T=140$~K, the shape and spectral position of the PL spectrum remain unchanged, and the exciton lifetime decreases by 6\% only when the magnetic field is increased from 0 to 3~T. Therefore, the magnetic field dependences for different temperatures unambiguously demonstrate that the exciton lifetime in CrSBr is sensitive to the magnetic order. 

Another important observation is related to the overall increase of exciton lifetime with temperature. The temperature dependences of $\tau_{\rm X}(T)$ for $B=0$ and $B>B_{\rm S}$ are shown in Fig.~\ref{fig3}(d). The exciton lifetime increases non-linearly with the temperature increase. Temperature activation behavior is observed in the AFM phase at $B=0$ (green circles in Fig.~\ref{fig3}(d)), in the FM phase for temperatures below $T_{\rm N}$ (red diamonds), and persists into the PM (triangles), where the lifetime is practically independent of the magnetic field.

\section{\textit{Ab initio} simulation of excitonic states}
\label{ab_initio}

\begin{figure}[h!]
    \centering
    \includegraphics[width=0.85\linewidth]{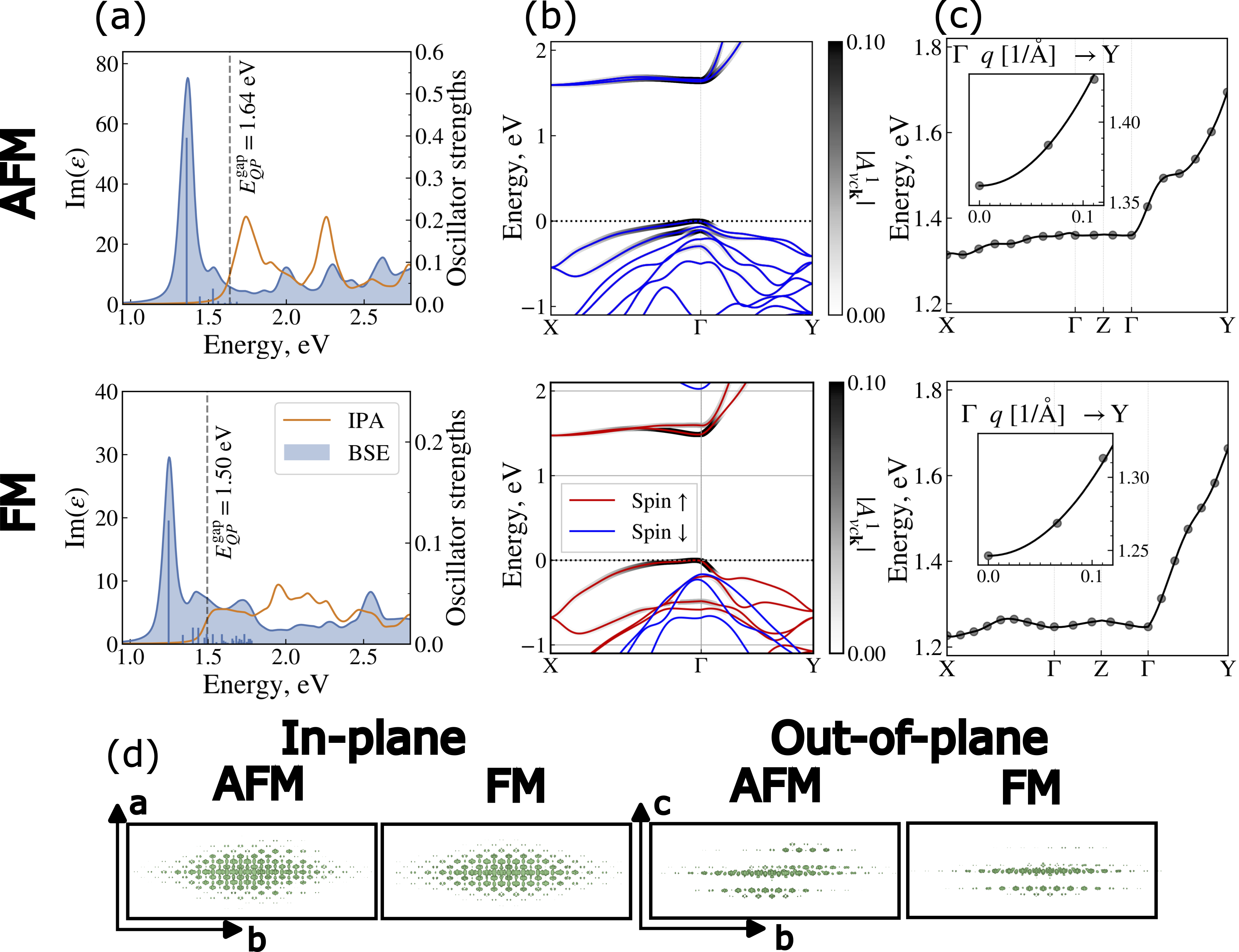}
    \caption{\textit{Ab initio} simulations of excitons in CrSBr. (a) Absorption spectra for light linearly polarized along the $b$-axis, obtained from the Bethe--Salpeter equation (BSE) and independent particle approximation (IPA), are shown for the AFM and FM phases. (b) Momentum-space exciton composition of the lowest energy exciton eigenstate projected on the IPA band structures. The amplitude of the single-particle states (see SI3, Eq.~(S2) for details) is shown by the black shadows under corresponding lines. (c) Dispersion curves for the lowest exciton, illustrating significant anisotropy in the transferred electron-hole momenta. The black line is meant to be a guide for the eye. Insets show the parabolas (solid lines) in the $\Gamma\rightarrow Y$-direction used to estimate the curvature of the exciton dispersion and mass. (d) Real-space electron probability density of the lowest excitonic state with the hole positioned near one of the sulfur atoms. Left panels display the in-plane profiles, where Cr–S chains extend along the $b$-axis. Right panels show the out-of-plane profiles.}
\label{fig:main_calc}
\end{figure} 

To gain microscopic insight into how the excitonic properties differ between AFM and FM phases, we have performed \textit{ab initio} simulations of both phases by numerically solving the Bethe–Salpeter equation (BSE), using inputs derived from the DFT$+U$$+J$ ground-state calculations (see Supplementary Information S3 for details of the calculation). The BSE absorption spectra for linearly polarized light in the crystallographic $b$-direction are shown in Fig.~\ref{fig:main_calc}(a) by blue color. The spectra obtained in the independent particle approximation (IPA) are shown (orange dashed lines) for comparison and demonstrate the importance of electron-hole interactions in shaping the optical response. The lowest exciton peak shows a redshift when transitioning from the AFM to the FM phase which is consistent with the experiment. From our BSE calculations, the oscillator strength of the lowest exciton is larger in the AFM phase than in the FM phase (blue bars in Fig.~\ref{fig:main_calc}(a)), which would correspond to a shorter radiative lifetime. We note that this trend is opposite to what is observed experimentally. This could hint at other recombination processes not captured by the BSE, e.g. localization of excitons.

Both the redshift and the weakening of the oscillator strength can be partially attributed to differences in the band structures and dominant optical transitions contributing to the excitons. In the AFM phase, the main single-particle states that contribute to the lowest energy exciton involve the three topmost valence bands and the two lowest conduction bands in a region around the $\Gamma$-point, as illustrated in Figure~\ref{fig:main_calc}(b), see SI3 for details. In the FM phase, the altered magnetic order modifies the exchange interaction, causing spin-dependent shifts and a reordering of the spin-polarized band structure near the $\Gamma$-point. This reconfiguration leads to an increased band separation between key valence and conduction bands that dominate the lowest exciton in the AFM phase, thereby weakening the corresponding optical transitions and reducing the oscillator strength of the lowest exciton in the FM phase by a factor of 3.2.

The observed redshift in the exciton peak when transitioning from AFM to FM phase, can be largely attributed to a reduction of the direct band gap, which implies a slightly larger dielectric screening. The calculated energy of the lowest exciton is $E^\text{AFM}_\text{exc,1} = 1.36$ eV in the AFM phase and $E^\text{FM}_\text{exc,1} = 1.25$ eV in the FM phase. The direct band gap decreases by $\Delta E_\text{gap} = 0.14$ eV in the FM phase, and the binding energy of the AFM exciton is 0.03 eV larger than the FM exciton.

To study the free exciton wavefunction in real space, we computed the real-space electron probability density for the lowest exciton in each magnetic phase, positioning the hole where the electron density of the valence bands was the highest at the $\Gamma$-point (around the S atom). As seen in Fig.~\ref{fig:main_calc}(d), the exciton displays strong anisotropy in both magnetic phases, with a large spatial extension in the $b$-direction and confinement along $a$- and $c$-directions. To quantify the in-plane anisotropy, we fitted the microscopic wave functions with the variational exciton wave function proposed in Ref.~\onlinecite{Semina2025}: $\Psi(x,y) \propto \exp\left(-\alpha\sqrt{x^2+\beta^2 y^2} \right)$. The resultant wave function shows high anisotropy with extent in $x$ direction $\alpha^{-1} \approx 5~$\AA\ and in $y$ direction $\alpha^{-1}\beta^{-1} \approx 20$~\AA\ in both phases within 5\% difference. Thus, the change of the oscillator strength is mostly related to the overlap of the Bloch parts of electron and hole wave functions and not to the overlap of their envelope functions  $\Psi(0)$. The anisotropy is also evident from the excitonic dispersion (Fig.~\ref{fig:main_calc}(c)), which was obtained by solving the BSE for the lowest exciton state at finite transferred electron-hole momenta. Both phases exhibit flat excitonic bands in the $\Gamma$--$X$-segment but dispersive ones in the $\Gamma$--$Y$-segment. The mass in the $\Gamma\to Y$-direction $M$ was estimated from the best fit of the energy of the exciton
\begin{equation}
    E_{\text{exc},1}(\Delta q_y) \approx  E_{\text{exc},1}(\mathbf{0}) + \frac{\hbar^2 (\Delta q_y)^2}{2M}, \quad \text{for } \Delta q_y \ll 1\ \text{\AA}^{-1},
    \label{eq:mass}
\end{equation}
where the energy $E_{\text{exc},1}(\Delta q_y)$ was calculated by solving BSE for small wave vectors $\Delta q_y$ along $\Gamma$--$Y$ path in $q$-space, see inset of Fig.~\ref{fig:main_calc}(c), and $\hbar$ is the Planck constant. Both magnetic phases exhibit similar curvatures close to the $\Gamma$-point, and a convergence study with decreasing $\Delta q_y$ suggests effective masses of approximately $M=0.73$$m_0$ for the FM phase and $M=0.67$$m_0$ for the AFM phase, where $m_0$ is the free electron mass (see Supplementary Information for more details).

\section{Discussion}

Our main experimental results are the switching of the exciton lifetime $\tau_X$ related to the transition between the AFM and FM phases of CrSBr and the significant increase of the exciton lifetime $\tau_X$ with temperature in both phases. Let us first exclude non-radiative exciton recombination and slow exciton energy relaxation as possible contributions to the observed magnetic switching of the PL transients.

At first glance, the magnetic field dependence of $\tau_X$ could be attributed to the increasing role of non-radiative process in the FM phase. Indeed, FM-ordering enables the transport of excitons along the $c$-axis, which is otherwise impossible in the AFM phase. Mobile excitons reach defect states or surface more quickly and are, therefore, more likely to recombine non-radiatively.  For such a process, an increase in temperature should lead to an enhancement of  non-radiative recombination, which in turn reduces the exciton lifetime.  However, an overall increase of $\tau_X$ with temperature increase in the experiment excludes such a scenario. Moreover, the time-integrated signal does not show significant changes with temperature increase and the PL yield remains approximately the same in the whole range up to 150~K. All these points indicate the negligible contribution of non-radiative recombination at low temperatures below 100~K.

Slow energy relaxation of excitons is important when the intrinsic radiative lifetime is significantly shorter than the time required for excitons to reach thermal equilibrium. In this case the energy relaxation time $\tau_\epsilon$ governs the flow of excitons into the radiative states with lower energy followed by their immediate recombination. The exponential decay of the PL signal is then determined by $\tau_\epsilon$, as demonstrated in MoSe$_2$ heterostructures~\cite{Fang-2019}. However, the exciton thermalization should be more efficient at higher temperatures. This would result in shortening of $\tau_\epsilon$ and consequently $\tau_X$, which is in contrast to our observation. Our data resemble rather a scenario in which an increase of temperature leads to a population of excitonic states outside the radiative cone, resulting in an extended exciton recombination time, as also reported for 2D vdW materials~\cite{Korn-2011, Robert-2016}. Moreover, recent study of exciton dynamics in  AFM phase of CrSBr reported energy relaxation within the first picosecond after optical excitation~\cite{Meineke-2024}, similar to conventional diluted magnetic semiconductors, such as CdMnTe~\cite{Kossut-book}. We, therefore, conclude that exciton thermalization occurs much faster than population decay, so that the observed PL dynamics is governed by the radiative recombination of excitons.

Changes in the exciton radiative lifetime can be attributed to variation in the oscillator strength. Indeed {\it ab initio} calculations in previous section show that the oscillator strength changes by the factor 3.2 when switching between the magnetic phases of CrSBr. However, according to these calculations, the opposite trend is expected: namely, a shorter $\tau_X$ in the AFM phase. BSE calculations demonstrate also that the excitons in CsSBr are quasi-1D in both phases: along the $\Gamma$--$X$ direction the effective mass is much larger, leading to confinement within a few atomic layers along the $a$- and $c$-axes, while the mass along the $b$-direction is relatively small ($0.73 m_0$ in FM and $0.67 m_0$ in AFM phase). For a one-dimensional free exciton, the radiative lifetime is expected~\cite{Citrin1992} to increase with temperature as $\tau_{\rm FX} \propto \sqrt{T}$ [blue dashed line in Fig.~\ref{fig3}(d)], in contrast to the much steeper experimental dependence. In addition, the maximum kinetic energy of excitons within the light cone $E_1= \hbar^2k_0^2/2M$ ($k_0$ is the light wave vector) is small as compared with characteristic temperatures when the increase of time is observed. Therefore, we propose that the PL emission originates from localized excitons, a scenario particularly relevant for systems with reduced dimensionality, where localization effects become more pronounced~\cite{Abrahams-1979}. The exciton localization volume increases upon transition from the AFM to FM phase due to reduced confinement along the $c$-axis. This results in an enhanced oscillator strength for the localized excitons and, consequently, a shorter $\tau_{X}$.

In order to treat the increase of the radiative time with the temperature quantitatively we adopt the concept of thermal population of dark exciton states out of the light cone, which leads to the increase of exciton radiative times with the temperature  in quantum wells \cite{Andreani1991, Andreani-book}. For one-dimensional excitons we employ the theory developed by Citrin~\cite{Citrin1992} and later applied by Akiyama et al.~\cite{Akiyama1994} to describe the temperature dependence of radiative lifetimes in quantum wires. Importantly, this model takes into account the exciton localization  at defects and their thermal activation in free exciton states \cite{Lomascolo1998, Cade2004}. The combined effects of localized and free excitons, together with the thermal population of dark exciton states outside the light cone, yield the following expression~\cite{Lomascolo1998}:
\begin{equation}\label{eq:tau_fit}
 \tau(T) = \frac{N_D \exp(E_{\rm loc}/k_BT) + \sqrt{\frac{2Mk_B}{\hbar^2\pi^2}}\sqrt{T} }{\frac{N_D}{\tau_{\rm loc}} \exp(E_{\rm loc}/k_BT) + \frac1{\tau_0}\sqrt{\frac{2M E_1}{\hbar^2\pi}}}\,,
\end{equation}
where 
$k_B$ is the Boltzmann constant, $M$ is the exciton mass defined in Eq.~\eqref{eq:mass},  $N_D$ is the effective density of localization centers along $Y$ direction ($b$-axis) with exciton localization energy $E_{\rm loc}$ and $\tau_{\rm loc}$ is the localized exciton decay time, $\tau_0$ is the intrinsic exciton decay time $\tau_0 =  nm_0c/(\pi e^2 f_X)$, where $n$ is the refractive index, $m_0$ is the free electron mass, $c$ is the velocity of the light in the vacuum, and $f_X$ is the oscillator strength per unit area, {\it ab initio} calculations give $f_X^{\rm AFM} \approx 3.96\times 10^{-3}$~\AA$^{-2}$ and $f_X^{\rm FM} \approx 1.22\times 10^{-3}$~\AA$^{-2}$) and $E_1= \hbar^2k_0^2/2M = E_X(n/c)^2/2M$ is the maximum kinetic energy of excitons which can decay radiatively. By using exciton energies in both phases $E_X^{\rm FM} = E_X^{\rm AFM} = 1.3$~eV, refractive index corresponding to the polarization of the active exciton transition $n=4.5$~\cite{Wang2023} and masses obtained from BSE, one obtains the following values in Eq.~\eqref{eq:tau_fit}: maximum kinetic energies in the light cone $E_1^{\rm AFM}= 38$~$\mu$eV, $E_1^{\rm FM}= 35$~$\mu$eV and the characteristic radiative time of free 1D exciton $\tau_{0,{\rm BSE}}^{\rm AFM}= 4.3$~ps, $\tau_{0,{\rm BSE}}^{\rm FM}= 14$~ps. 
These numbers are in reasonable agreement with the parameters obtained by the direct fit of experimental data given in Table~\ref{tab:fit}.

\begin{table}[t]
\centering
\caption{Fit parameters  in Eq.~(\ref{eq:tau_fit}) describing exciton localization in AFM and FM phases of CrSBr obtained for the data in Fig.~\ref{fig3}(d).}
\label{tab:fit}
\begin{tabular}{lcccc}
\toprule
Phase & $N_D$ (cm$^{-1}$) & $E_{\mathrm{loc}}$ (meV) & $\tau_{\mathrm{loc}}$ (ps) & $\tau_0$(ps) \\
\hline
AFM & $0.5\times10^{4}$ & 40 & 11.7 & 10\\
FM  & $2\times10^{4}$ & 32 & 7.5 & 32\\
\hline
\end{tabular}
\end{table}

To compare with the experimental data, in Eq.~(\ref{eq:tau_fit}) we assume that excitons possess larger localization energies in the AFM phase compared to the FM phase, i.e. $E_{\rm loc}^{\rm AFM} > E_{\rm loc}^{\rm FM}$, which consequently leads to $\tau_{\rm loc}^{\rm AFM} > \tau_{\rm loc}^{\rm FM}$. The effective density of localization sites also changes during the AFM--FM transition. We expect that, due to the reduced confinement between the vdW layers, the effective distance in $Z$ direction ($c$-axis)  accessible for exciton increases and the density becomes larger in the FM phase, i.e. $N_D^{\rm AFM} < N_D^{\rm FM}$. 
The reasonable fit of the data with Eq.~(\ref{eq:tau_fit}) can be obtained for $\tau_0 \geq 10$~ps. Otherwise, the temperature dependence of the lifetime of free excitons $\tau_{\rm FX}=\sqrt{\pi k_B T / E_1} \tau_0$, which is shown for the AFM phase with the dashed blue curve in Fig.~\ref{fig3}(d), contradicts the experimental data, since at high temperatures the measured lifetimes are limited from above by $\tau_{\rm FX}$. For this reason, we obtain  slightly larger $\tau_{0}^{\rm AFM}=10$~ps and $\tau_{0}^{\rm FM}=32$~ps compared to the BSE calculations, while preserving the relation between the time constants in the AFM and FM phases. The fitting parameters are summarized in Table~\ref{tab:fit}, and the corresponding curves are shown in Fig.~\ref{fig3}(d). Note that the resulting curves do not reproduce the data in the paramagnetic phase for $T>132$~K. Because of the large number of fitting parameters, a reliable analysis of the paramagnetic phase is difficult and requires further studies in a wider temperature range. Nevertheless, the observed temperature dependence of the exciton lifetime in the FM and AFM phases suggests that exciton localization accounts for the experimentally observed magnetic switching of the exciton lifetime. 

\section{Conclusion}

To summarize, we demonstrate strong reduction of the exciton lifetime $\tau_{\rm X}$ from 11 to 7~ps during the transition from the AFM to FM phase at low temperature of 2~K. We attribute this behavior to variations in the radiative lifetime of excitons caused by changes in the exciton localization volume, which increases due to reduced confinement between the vdW layers upon the AFM--FM transition. This interpretation is supported by a pronounced increase of $\tau_{\rm X}$ with temperature, allowing us to estimate localization energies in the range of $30-40$~meV. We emphasize that localization effects at low temperatures play a crucial role and should be carefully considered in future studies of CrSBr layers.

\section{Methods}

\textbf{Sample preparation.} 

CrSBr single crystals were grown by the chemical vapor transport technique. Stoichiometric quantities of Cr powder (99.0\%, Goodfellow), S powder (99.99\%, Sigma Aldrich)  and liquid Br (99.8\%, Thermofisher) are inserted into a quartz tube with a small excess (5\%) of Br to act as the transport agent and the tube is sealed under vacuum. The mixture is first pre-reacted by slowly increasing the temperature up to 700$^{\circ}$C over two weeks. The tube is then placed in a two-zone furnace with the source zone heated to 700$^{\circ}$C and the growth zone to 900$^{\circ}$C and maintained at this temperature for 48~hrs. The source zone is then heated to 900$^{\circ}$C while the growth zone is held at 850$^{\circ}$C. Following this, the source zone is slowly heated to 940$^{\circ}$C over a two-week period, while the growth zone is cooled to 800$^{\circ}$C during the same period. After this the furnace is cooled to room temperature and the tubes were removed.

CrSBr flakes were mechanically exfoliated on a 290-nm SiO$_2$/Si substrate with a low-tack, low-residue tape (Nitto BT-130E-SL tape) from bulk crystals. Before exfoliation, the substrate was cleaned by acetone (10 min) and isopropanol (10 min) with an ultrasonic bath and further treated with an oxygen plasma to remove contaminants and residues.

\textbf{Atomic force microscopy.} The atomic force microscopy measurements were performed in a Bruker Dimension Edge in a tapping mode.

\textbf{Time-integrated and time-resolved photoluminescence.}
The sample was mounted on a three-axis piezo stage within a variable temperature insert (VTI) of a magneto-optical bath cryostat with  a split-coil superconducting magnet, capable of generating magnetic fields up to 6~T. In most experiments, except for those exploring temperature dependence, the sample was immersed in superfluid helium at approximately 2~K. Laser excitation and photoluminescence (PL) detection were performed in a confocal geometry along the $c$-axis of the flake using an achromatic lens with a numerical aperture of 0.5  and a focal length of 10~mm, located inside the VTI of the cryostat.

A tunable pulsed laser with a pulse duration of 100~fs and a repetition rate of 80~MHz was used for optical excitation. The photon energy was adjusted to 1.771~eV, with approximately 1~mW of power used for time-integrated measurements, while lower power levels were employed for time-resolved measurements. The laser was focused into a spot approximately 3~$\mu$m in diameter. The magnetic field was applied in Voigt or Faraday geometries, aligning with one of the crystallographic axes ($a$ or $b$ in Voigt geometry and the $c$-axis in Faraday geometry).

The collected PL signal was dispersed in a single-stage spectrometer with a linear dispersion of $\sim 3.0$~nm/mm with 600~g/mm grating. Time-integrated data were acquired using a nitrogen-cooled charged-coupled device (CCD) camera. For time-resolved measurements, the spectrometer was set to zero order, and two filters (shortpass and longpass) with resulting bandpass interval 850-1000~nm were used. A tunable interference filter with a bandwidth of 7~nm was also used to examine the wavelength dependence. Linear polarization for excitation and detection is achieved using Glan-Thompson prisms in conjunction with half-wave plates.




\textbf{ACKNOWLEDGMENTS}
We are grateful to M.M.~Glazov, D.R.~Yakovlev, and V.L.~Korenev for valuable discussions. I.V.K. acknowledges  Deutsche Forschungsgemeinschaft via project KA 6253/1-1, no. 534406322, M.O.N. acknowledges Deutsche Forschungsgemeinschaft within the SPP 2196 (project no. 506623857). The work in Dortmund was supported by the Deutsche
Forschungsgemeinschaft through the Collaborative Research Center TRR 142/3 (Grant No. 231447078, Projects A02 and A10). The work is supported by the European Union’s Horizon Europe research and innovation program under the Marie Sklodowska-Curie grant agreement 101118915 (TIMES) and it is part of the project I+D+i PID2023-146181OB-I00 UTOPIA, funded by MCIN/AEI/10.13039/501100011033, and the project PROMETEO/2024/4 (EXODOS). T.G. and A.M.S. acknowledge computer resources at Agustina and technical support provided by BIFI and Barcelona Supercomputing Center (FI-2025-2-0001), and the computer resources at Tirant-UV (project lv48 - FI-2025-2-0001). 
Z.S. is supported by project LUAUS25268 from Ministry of Education Youth and Sports (MEYS) and by the project Advanced Functional Nanorobots (reg. No. CZ.02.1.01/0.0/0.0/15\_003/0000444 financed by the EFRR). Z.S. acknowledge the assistance provided by the Advanced Multiscale Materials for Key Enabling Technologies project, supported by the Ministry of Education, Youth, and Sports of the Czech Republic. Project No. CZ.02.01.01/00/22\_008/0004558, Co-funded by the European Union.
A.I.T., Y.W., X.H. and D.J.G. thank EPSRC grants EP/V006975/1, EP/V007696/1, EP/V026496/1, EP/S030751/1.
Y.W. and A.I.T. acknowledge support from the UKRI fellowship TWIST-NANOSPEC EP/X02153X/1. The crystal growth at Warwick was supported by two Engineering and Physical Sciences Research Council grants: Grant No. EP/T005963/1 and the U.K. Skyrmion Project Grant No. EP/N032128/1.

%

%

\clearpage
\newpage
\onecolumngrid

\begin{center}
  \textsf{\textbf{\Large Supplementary Information:}}\\[0.2cm]
  \textbf{\large {\thetitle}}
\end{center}

\hypersetup{pageanchor=false}  

\setcounter{equation}{0}
\setcounter{figure}{0}
\setcounter{table}{0}
\setcounter{section}{0}
\setcounter{page}{1}
\renewcommand{\thepage}{S\arabic{page}}
\renewcommand{\theequation}{S\arabic{equation}}
\renewcommand{\thefigure}{S\arabic{figure}}
\renewcommand{\thetable}{S\arabic{table}}
\renewcommand{\thesection}{S\arabic{section}}
\renewcommand{\theHsection}{Ssection.\arabic{section}}
\renewcommand{\thesubsection}{} 
\renewcommand{\bibnumfmt}[1]{[S#1]}
\renewcommand{\citenumfont}[1]{S#1}
\renewcommand{\theHequation}{Sequation.\arabic{equation}}  
\renewcommand\theHfigure{Sfigure.\arabic{figure}}  



\section{Atomic force microscopy data}\label{SI:AFM}

\begin{figure}[h!]
	\centering
	\includegraphics[width=1\linewidth]{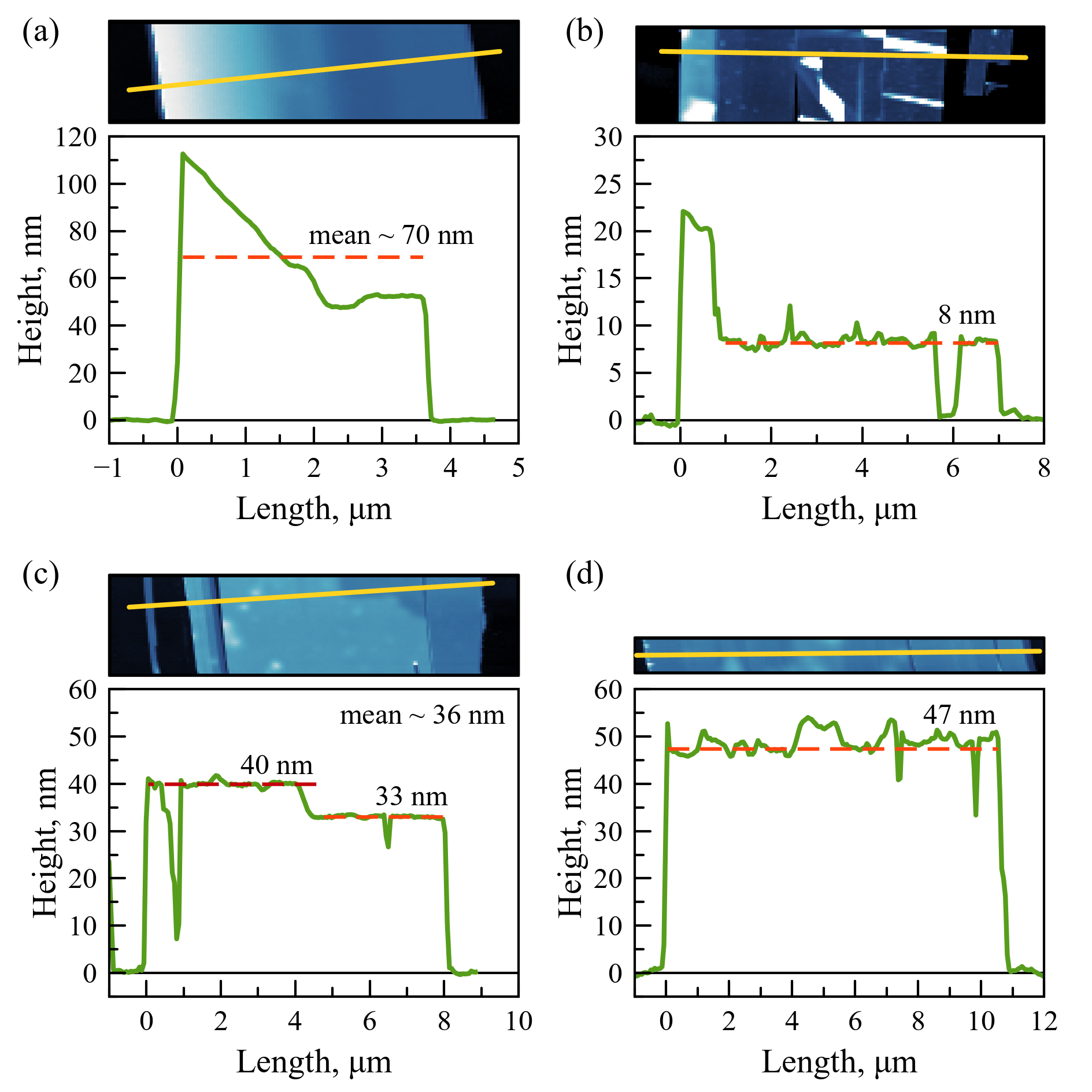}
	\caption{Exemplary atomic force microscopy data for different CrSBr flakes: (a) with average thickness of $\sim$70~nm, (b) 8~nm, (c) $\sim$36~nm and (d) 47~nm. For each flake normalized color map is shown. Whitish areas are folds of the flakes (b) and bubbles under them (c, d), so that these areas do not describe real thickness. For the flake (a) the average value may be exaggerated due to possible lifting of the left side of the flake (area 0-2 $\mu$m) from the substrate. Assuming this lift, the lower limit of the flake (a) thickness is 51~nm. Yellow lines on the maps show the coordinates of the characteristic cuts plotted below. For comparison, the diameter of the laser beam is 3~$\mu$m. That means for some flakes, like one shown in (c), certain thickness cannot be optically probed, because laser excites areas with different thicknesses simultaneously. For this reason, in the text flakes are named after approximate average thicknesses of the optically probed areas. However, we do not observe correlation of the exciton lifetime on the flake thickness in the measured range.
    }
	\label{fig-AFM}
\end{figure}

The atomic force microscopy measurements are shown in Figure~\ref{fig-AFM} for several characteristic CrSBr flakes. Measurements shown in the main text were perform on flakes (a), (b) and (c) and referred to as $\sim$70~nm, 8~nm and $\sim$36~nm according to their thickness.

\section{Reflectivity and magnetic field induced ellipticity spectra} \label{SI:MCD}

\begin{figure}[h!]
	\centering
	\includegraphics[width=1\linewidth]{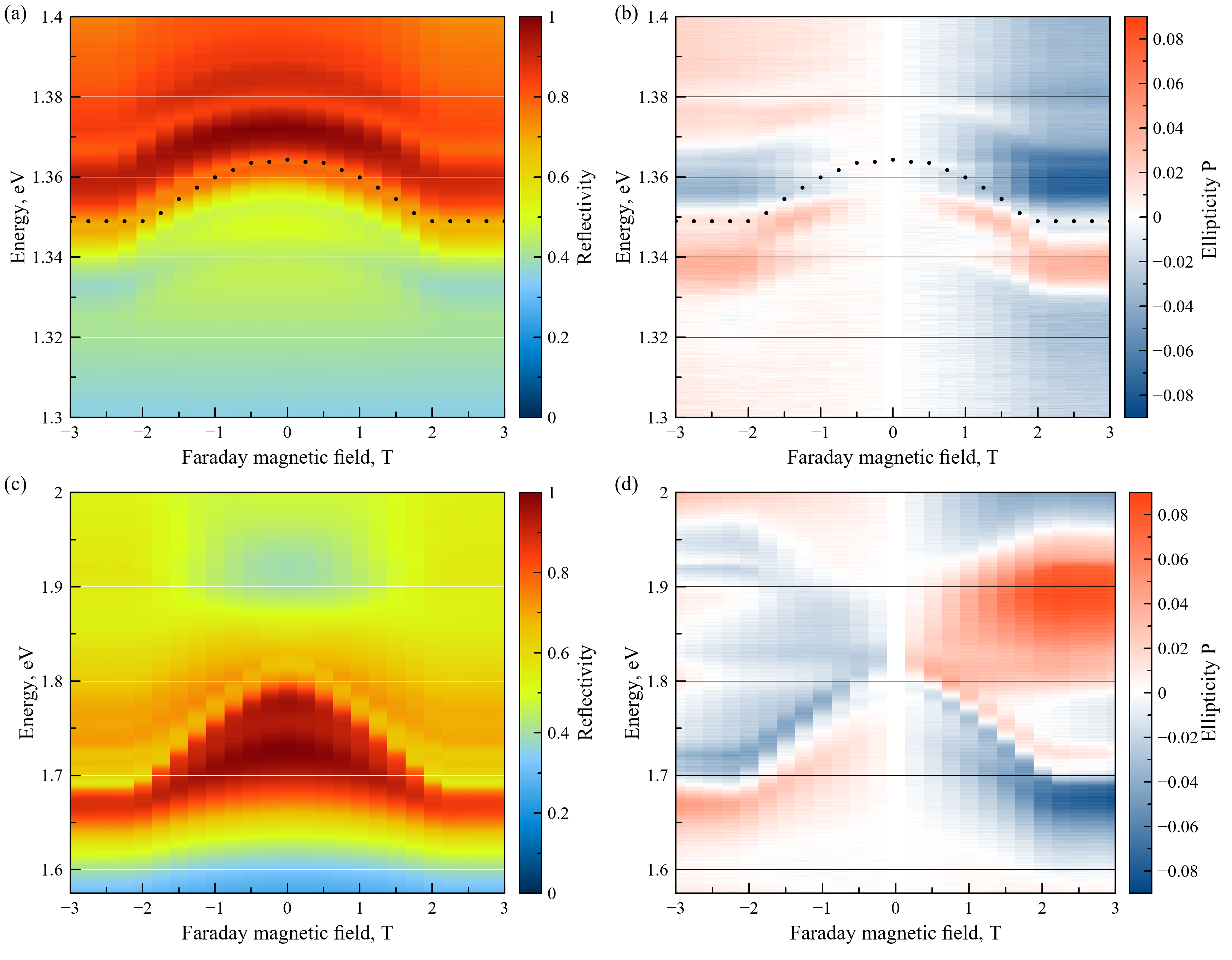}
	\caption{Reflectivity measurements. Excitation is modulated $\sigma^+/\sigma^-$ white light, detection in linear polarization parallel to the $b$-axis. Panels (a, b) and (c, d) show two detection ranges. (a) and (c) are the color maps of total reflectivity, while (b) and (d) are ellipticity signals (circular polarization defined as relative difference between $\sigma^+$ and $\sigma^-$ reflectivity spectra).  The polarization $P = \frac{R^+ - R^-}{R^+ + R^-}$, where $R^{+(-)}$ is the reflectivity signal under $\sigma^{+(-)}$-polarized white light. Additionally, the data are normalized to the zero polarization in zero magnetic field by the introduction of the coefficient $\alpha(E)$ for $R^-$ component. Black dots in (a, b) shows the position of 1s exciton evaluated from PL spectra.}
	\label{fig-MCD}
\end{figure}

Figure~\ref{fig-MCD} summarizes the reflectivity measurements performed in the Faraday geometry ($\mathbf{B}\|c$) using a broadband white-light source and spectrometer for detection. Two types of measurements are shown: (i) Reflectivity spectra $R$ at different magnetic fields in linear polarization along $b$-axis; (ii) Magnetic field induced ellipticity ($P = \frac{R^+ - R^-}{R^+ + R^-}$), where $R^{\pm}$ are the reflectivity spectra measured for circularly polarized incident light $\sigma^\pm$, respectively. Panels (a) and (b) show the results for the ground exciton state, for $R$ and $P$, respectively. A redshift of about 16~meV is observed in correspondence with the PL data. The ellipticity signal for the ground state is even with respect to the magnetic field, indicating that no splitting of the exciton states occurs in this case. In contrast, a markedly different behavior is observed for the excited states, shown in panels (c) and (d). First, a very large energy shift of approximately 100~meV is detected. Second, the ellipticity signal becomes odd with respect to the magnetic field, demonstrating that the corresponding bands split with opposite signs depending on the field direction.

\clearpage
\section{\textit{Ab initio} computational details}
\label{SI:ab_initio}
\subsection{DFT calculations}
The ground state calculations were performed with the Quantum ESPRESSO DFT code (v7.3.1) \cite{QE-2017}  using the PBE exchange-correlation functional in combination with scalar relativistic optimized norm-conserving Vanderbilt pseudopotentials \cite{hamann2013optimized}. A plane wave cutoff of 120 Ry and a charge density cutoff of 480 Ry were used. In the bulk FM case, a $\Gamma$-centered Monkhorst-Pack mesh of size $12\times 10\times 5$ was used to sample the Brillouin zone. The bulk AFM configuration was obtained by extending the FM unit cell along the stacking direction and imposing antiparallel magnetic order between successive layers. Consequently, a smaller mesh of $12\times 10\times 3$ proved sufficient for well-converged values of the total energy and band gap. The spin degree of freedom was treated in a collinear fashion without spin-orbit coupling. Spin-orbit coupling was tested and found to have negligible impact on the position and energy of the first excitonic peak of the BSE absorption spectrum and was therefore omitted to prioritize convergence of other computational parameters. The Grimme-D2 correction for the van der Waals interactions was used in all DFT calculations. Both the atomic positions and orthorhombic lattice parameters were relaxed for the FM and AFM bulk structures using the BFGS algorithm until the forces on all of the atoms were below $10^{-3}$ Ry/Bohr, the total energy between two consecutive BFGS steps was less than $10^{-4}$ Ry, and the residual pressure was less than 0.5 kbar. The structural relaxation was carried out without Hubbard corrections, as atomic forces with an intra-atomic exchange constant $J$ are not supported in the version of Quantum ESPRESSO used in this work. The relaxed structure preserved the orthorhombic symmetry corresponding to the point group D$_{2\text{h}}\, (mmm)$, yielding lattice parameters of $a=3.52$ Å, $b=4.73$ Å and $c=8.05$ Å for both magnetic phases, which deviate $\sim 1\%$ from the experimental values reported for bulk AFM CrSBr \cite{telford2020layered}. Including a Hubbard parameter $U=4.0$ eV yielded a deviation $<2.5$\% compared to the experimental ones, supporting the use of the Hubbard-free relaxed geometry in subsequent BSE calculations.

The Hubbard parameters $U=2.5$ eV and $J=1.2$ eV were chosen to best reproduce the qualitative features of the band structure of the ferromagnetic monolayer obtained using the HSE06 hybrid functional \cite{wang2020electrically}. In particular, the overall shape of the bands near the Fermi level was well reproduced, although the DFT$+U$$+J$ band gap (1.29 eV) was smaller than the HSE06 one (1.66 eV). This underestimated gap, also observed in bulk calculations, was subsequently corrected using a rigid scissor shift in the BSE calculations. Additionally, these values were selected to minimize the conduction band splitting at the $\Gamma$-point in the monolayer and AFM bulk, which low-temperature PL measurements suggest is $\sim 35$ meV across different thicknesses \cite{Lin-2024b}. While this interpretation is based on peaks in the absorption spectra and may not directly reflect the quasiparticle band splitting, it provides a useful experimental constraint for tuning the Hubbard parameters. We note that the precise quasiparticle band structure of CrSBr remains a subject of ongoing investigation, and our choice of parameters reflects a balance between different theoretical and experimental constraints available at the moment.

\subsection{BSE calculations}
Bethe–Salpeter equation (BSE) calculations for both magnetic phases were carried out with the Yambo code (v5.3) \cite{sangalli2019many} , using input derived from the DFT$+U$$+J$ ground-state electronic structure as described above. A rigid scissor of 0.32 eV was applied to both phases and spin channels to reproduce the first excitonic peak for the AFM phase at 1.36 eV, as observed experimentally. This places the quasiparticle gap at 1.64 eV for the AFM phase, in good agreement with previously reported experimental and computational values, which range from approximately 1.5 eV to 2.1 eV   \cite{telford2020layered,Klein-2023,heissenbuttel2025quadratic,bianchi2023paramagnetic}. 

Since the DFT$+U$$+J$ band gap for the FM phase was 0.14 eV smaller than that of the AFM phase, and the same scissor shift was applied, the resulting quasiparticle gap for the FM configuration was 1.50 eV. The lower quasiparticle band gap in the FM phase compared to the AFM phase has also been observed in previous $GW$-BSE studies \cite{heissenbuttel2025quadratic}, suggesting that this trend is robust across different levels of theory. The number of bands per spin channel used to construct the non-interacting density response function was 300 for the AFM phase and 150 for the FM phase. The dielectric energy cutoff was set to 3 Ry, and the Brillouin zone was sampled with the same meshes as in the ground state calculations. Six conduction and six valence bands per spin channel were used to build the BSE kernel for both FM and AFM phases. The BSE Hamiltonian was solved using an inversion solver to obtain the frequency-dependent absorption spectrum up to 4.0 eV. Since this method does not provide excitonic eigenvectors, a separate calculation was performed using a recursive algorithm based on the SLEPc library \cite{hernandez2005slepc} to compute the 200 lowest excitonic states and construct the corresponding real-space exciton wavefunctions. 

In order to analyze the oscillator strengths of the low-energy excitons, we made use of the following expression for the diagonal elements of the macroscopic dielectric tensor \cite{fuchs2008efficient}:
\begin{equation}
    \varepsilon_{jj}(\omega) = 1 + \frac{e^2\hbar^2}{\Omega\varepsilon_0m_0}\sum_\lambda\frac{f_{jj}^\lambda}{E_\lambda}\sum_{\beta=\pm 1}\frac{1}{E_\lambda-\beta\hbar(\omega+i\gamma)}
\end{equation}
where the dimensionless oscillator strengths are defined with
\begin{equation}
    f_{jj}^\lambda = \frac{m_0E_\lambda}{\hbar^2N_k}\left|\sum_{vc\mathbf{k}m}\bra{c\mathbf{k}m}r_j\ket{v\mathbf{k}m}^*A^\lambda_{vc\mathbf{k}m}\right|^2.
\end{equation}
Here $E_\lambda$ denotes the exciton energies, $r_j$ is the $j$-th Cartesian component of the position operator, $A^{\lambda}_{vc\mathbf{k}m}$ is the eigenvalue component in transition basis between valence band $v$ and conduction band $c$ at k-point $\mathbf{k}$, and spin $m$ (collinear setup with no spin flip transitions allowed). $N_k$ is the total number of k-points in the BZ, $m_0$ is the electron mass, and $\Omega$ is the unit cell volume. Each exciton eigenvector is normalized such that $1=\sum_{vc\mathbf{k}m}|A^\lambda_{vc\mathbf{k}m}|^2$. Given that the AFM phase computations made use of an enlarged computational cell to accommodate the magnetic ordering, the AFM absorption spectrum was scaled by the ratio of the AFM and FM computational cell volumes to render both datasets comparable at the primitive-cell volume. The oscillator strengths for the lowest 200 eigenstates are displayed in Fig.~\ref{fig:main_calc} in the main text.

To characterize the localization and anisotropy of the exciton wavefunction we used the real space exciton density $\rho(\mathbf{r}_e,\mathbf{r}_h)$ by fixing the position of the hole $\mathbf{r}_h$ and projecting the exciton wavefunction onto the Bloch functions:
\begin{equation}
    \Psi_\lambda(\mathbf{r}_e,\mathbf{r}_h) = \sum_{vc\mathbf{k}m}A^\lambda_{vc\mathbf{k}m}\psi^*_{v\mathbf{k}}(\mathbf{r}_e)\psi_{c\mathbf{k}}(\mathbf{r}_h).
\end{equation}
The decay of $\rho$ around $\mathbf{r}_h$ was described by an anisotropic exponential form
\begin{equation}
    \rho(\mathbf{r}_e,\mathbf{r}_h,\alpha,\beta) = B\exp\left(-2\alpha\sqrt{(x_e-x_h)^2+\beta^2(y_e-y_h)^2}\right)
    \label{eq:alpha_beta}
\end{equation}
where $\alpha$ characterizes the overall localization and $\beta$ describes the in-plane anisotropy with $\beta<1$ corresponding to an elongation in the $y$-direction \cite{Semina2025}. Since $\rho$ is strongly peaked around $\mathbf{r}_h$, the fitting was performed along each of the principal directions by setting $y_e=y_h$ and $x_e=x_h$, in order to extract $\alpha$ and $\alpha\beta$, respectively. The fitted values can be seen in Table \ref{table:alpha_beta} for the hole positioned near a Cr atom and near a S atom.
\begin{table}[t]
    \centering
    \begin{tabular}{c c c c c}
    Hole site & $\alpha$ (AFM) & $\beta$ (AFM) & $\alpha$ (FM) & $\beta$ (FM)  \\  
    \hline
    \hline
    Cr & 0.21 Å$^{-1}$ & 0.33 & 0.19 Å$^{-1}$ & 0.32 \\ 
    S  & 0.20 Å$^{-1}$ & 0.26 & 0.21 Å$^{-1}$ & 0.26 \\ 
    \end{tabular}
    \caption{The fitted parameters for the exciton density according to Eq. \eqref{eq:alpha_beta}. }
    \label{table:alpha_beta}
\end{table}

The BSE was also solved for finite electron-hole transferred momenta $\mathbf{q}$ to access the dispersion for the lowest exciton and to extract effective masses around the $\Gamma$-point. The mass in the $\Gamma\to Y$-direction was calculated according to 
\begin{equation}
    E_{\text{exc},1}(\Delta q_y) \approx  E_{\text{exc},1}(\mathbf{0}) + \frac{\hbar^2 (\Delta q_y)^2}{2m^{\Gamma \to Y}_{\text{exc}}}, \quad \text{for } \Delta q_y \ll 1\ \text{\AA}^{-1}
    \label{eq:exc_mass}
\end{equation}
where $m^{\Gamma\to Y}_\text{exc}$ (denoted by $M$ in the main text) was extracted using the second-order central finite difference approximation $[E_{\text{exc},1}(-\Delta q_y) -2E_{\text{exc},1}(\mathbf{0}) + E_{\text{exc},1}(\Delta q_y)]/(\Delta q_y)^2$ for the second derivative with respect to $\Delta q_y$, utilizing the symmetry relation $E_{\text{exc},1}(\Delta q_y) = E_{\text{exc},1}(-\Delta q_y)$. Since the relation \eqref{eq:exc_mass} is assumed to hold for small $\Delta q_y$ values, we combined calculations from two Brillouin zone samplings with in-planar dimensions $16\times 12$ and $12\times 20$ for both magnetic phases, with the same out-of-plane sampling as before (the different sampling led to a relative change of the first exciton energy of less than 0.14\% for both phases). Consequently, we got a finer sampling of momenta points closer to the $\Gamma$-point in the $\Gamma\to Y$-direction, which allowed us to study the convergence of the exciton mass with respect to $\Delta q_y$ (Table \ref{table:exc_masses}).
\begin{table}[t]
    \centering
    \begin{tabular}{c c c}
    \hline
    $\Delta q_y$ (Å$^{-1}$) & FM $m_{exc}^{\Gamma\rightarrow Y}$ ($m_0$) & AFM $m_{exc}^{\Gamma\rightarrow Y}$ ($m_0$)  \\
    \hline
    0.066 & 0.728 & 0.673 \\
    0.110 & 0.709 & 0.700 \\
    0.133 & 0.969 & 0.980 \\
    0.221 & 1.215 & 1.399 \\ 
    \hline
    \end{tabular}
    \caption{The calculated exciton mass along $\Gamma\to Y$ using a finite difference approximation with step size $\Delta q_y$ for the second derivative of the energy with respect to the momentum.}
    \label{table:exc_masses}
\end{table}

\begin{figure}[b!]
    \begin{center}
        \makebox[\textwidth][c]{%
            \includegraphics[width=0.9\linewidth]{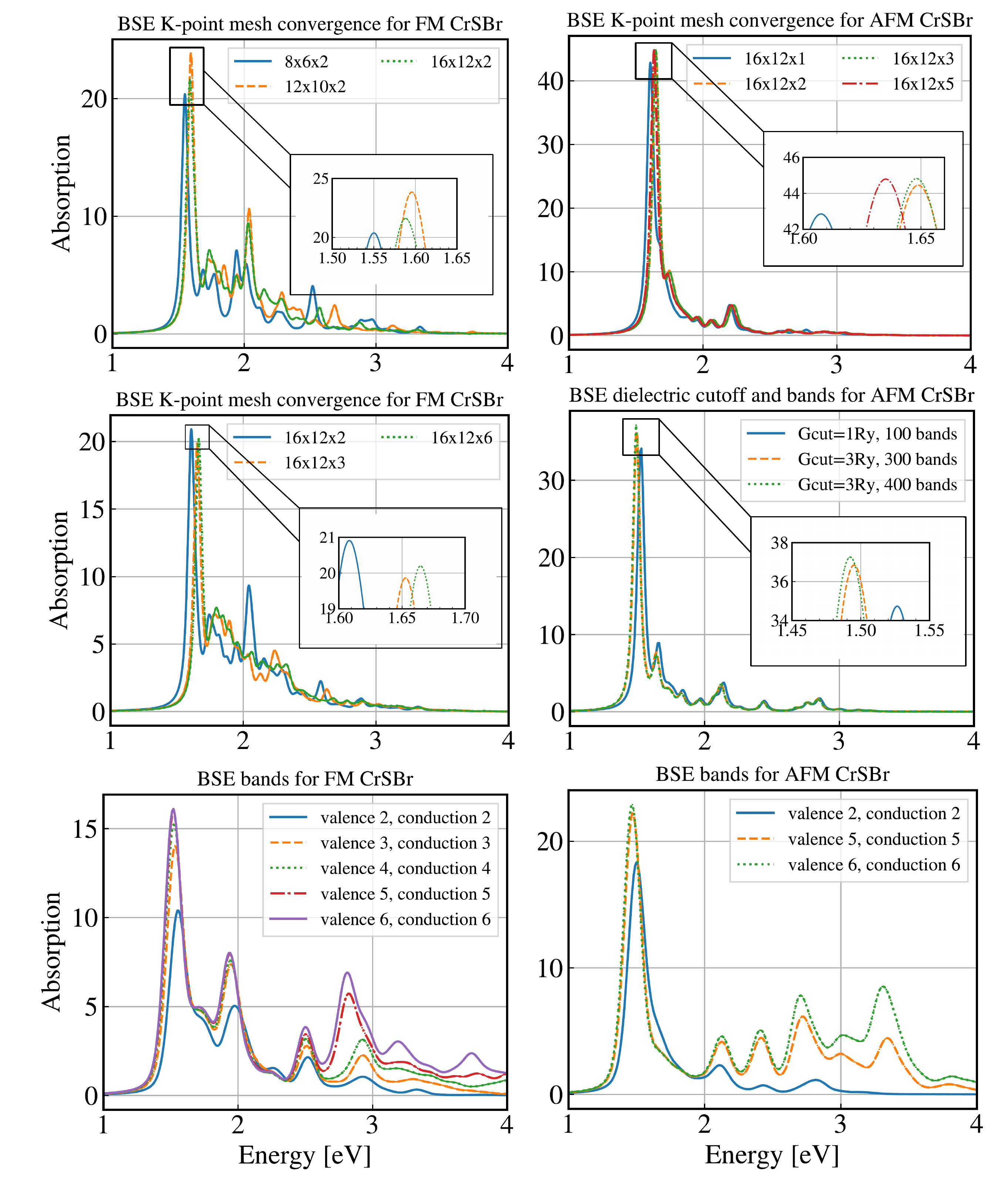}
        }
    \end{center}
    \caption{Convergence for the absorption along the crystallographic $b$-direction obtained with BSE calculations. Unless otherwise stated, the default parameters used were: an $8\times6\times2$ k-point mesh; a BSE kernel constructed with two valence and two conduction bands per spin channel; and a dielectric screening cutoff of 1 Ry using 100 bands for the FM phase and 200 bands for the AFM phase.}
    \label{fig:main_conv}
\end{figure}

The convergence of the main numerical parameters for the BSE calculation is shown in Figure~\ref{fig:main_conv}. We focused on converging the first exciton peak, as its position and intensity were the key observables that were of interest. In order to save computational costs, each parameter was individually converged while the others were held at reasonable but not fully converged values (see figure caption for details). Since the BSE results were sensitive to the underlying k-mesh, we additionally performed two calculations for the FM phase with k-meshes $16\times 12\times 5$ and $12\times 20\times 5$, and for the AFM phase $16\times 12\times 3$ and $12\times 20\times 3$, and found that the variations in the first peak positions and intensities were minor and well within an acceptable range.

\clearpage

\clearpage


%


\begin{thebibliography}{32}%
\makeatletter
\providecommand \@ifxundefined [1]{%
 \@ifx{#1\undefined}
}%
\providecommand \@ifnum [1]{%
 \ifnum #1\expandafter \@firstoftwo
 \else \expandafter \@secondoftwo
 \fi
}%
\providecommand \@ifx [1]{%
 \ifx #1\expandafter \@firstoftwo
 \else \expandafter \@secondoftwo
 \fi
}%
\providecommand \natexlab [1]{#1}%
\providecommand \enquote  [1]{``#1''}%
\providecommand \bibnamefont  [1]{#1}%
\providecommand \bibfnamefont [1]{#1}%
\providecommand \citenamefont [1]{#1}%
\providecommand \href@noop [0]{\@secondoftwo}%
\providecommand \href [0]{\begingroup \@sanitize@url \@href}%
\providecommand \@href[1]{\@@startlink{#1}\@@href}%
\providecommand \@@href[1]{\endgroup#1\@@endlink}%
\providecommand \@sanitize@url [0]{\catcode `\\12\catcode `\$12\catcode
  `\&12\catcode `\#12\catcode `\^12\catcode `\_12\catcode `\%12\relax}%
\providecommand \@@startlink[1]{}%
\providecommand \@@endlink[0]{}%
\providecommand \url  [0]{\begingroup\@sanitize@url \@url }%
\providecommand \@url [1]{\endgroup\@href {#1}{\urlprefix }}%
\providecommand \urlprefix  [0]{URL }%
\providecommand \Eprint [0]{\href }%
\providecommand \doibase [0]{https://doi.org/}%
\providecommand \selectlanguage [0]{\@gobble}%
\providecommand \bibinfo  [0]{\@secondoftwo}%
\providecommand \bibfield  [0]{\@secondoftwo}%
\providecommand \translation [1]{[#1]}%
\providecommand \BibitemOpen [0]{}%
\providecommand \bibitemStop [0]{}%
\providecommand \bibitemNoStop [0]{.\EOS\space}%
\providecommand \EOS [0]{\spacefactor3000\relax}%
\providecommand \BibitemShut  [1]{\csname bibitem#1\endcsname}%
\let\auto@bib@innerbib\@empty
\bibitem [{\citenamefont {Novoselov}\ \emph {et~al.}(2016)\citenamefont
  {Novoselov}, \citenamefont {Mishchenko}, \citenamefont {Carvalho},\ and\
  \citenamefont {Neto}}]{Novoselov-review2016}%
  \BibitemOpen
  \bibfield  {author} {\bibinfo {author} {\bibfnamefont {K.~S.}\ \bibnamefont
  {Novoselov}}, \bibinfo {author} {\bibfnamefont {A.}~\bibnamefont
  {Mishchenko}}, \bibinfo {author} {\bibfnamefont {A.}~\bibnamefont
  {Carvalho}},\ and\ \bibinfo {author} {\bibfnamefont {A.~H.~C.}\ \bibnamefont
  {Neto}},\ }\bibfield  {title} {\enquote {\bibinfo {title} {{2D} materials and
  van der {W}aals heterostructures},}\ }\href
  {https://doi.org/10.1126/science.aac9439} {\bibfield  {journal} {\bibinfo
  {journal} {Science}\ }\textbf {\bibinfo {volume} {353}},\ \bibinfo {pages}
  {aac9439} (\bibinfo {year} {2016})}\BibitemShut {NoStop}%
\bibitem [{\citenamefont {Wang}\ \emph {et~al.}(2022)\citenamefont {Wang},
  \citenamefont {Bedoya-Pinto}, \citenamefont {Blei}, \citenamefont {Dismukes},
  \citenamefont {Hamo}, \citenamefont {Jenkins}, \citenamefont {Koperski},
  \citenamefont {Liu}, \citenamefont {Sun}, \citenamefont {Telford},
  \citenamefont {Kim}, \citenamefont {Augustin}, \citenamefont {Vool},
  \citenamefont {Yin}, \citenamefont {Li}, \citenamefont {Falin}, \citenamefont
  {Dean}, \citenamefont {Casanova}, \citenamefont {Evans}, \citenamefont
  {Chshiev}, \citenamefont {Mishchenko}, \citenamefont {Petrovic},
  \citenamefont {He}, \citenamefont {Zhao}, \citenamefont {Tsen}, \citenamefont
  {Gerardot}, \citenamefont {Brotons-Gisbert}, \citenamefont {Guguchia},
  \citenamefont {Roy}, \citenamefont {Tongay}, \citenamefont {Wang},
  \citenamefont {Hasan}, \citenamefont {Wrachtrup}, \citenamefont {Yacoby},
  \citenamefont {Fert}, \citenamefont {Parkin}, \citenamefont {Novoselov},
  \citenamefont {Dai}, \citenamefont {Balicas},\ and\ \citenamefont
  {Santos}}]{Wang-review2022}%
  \BibitemOpen
  \bibfield  {author} {\bibinfo {author} {\bibfnamefont {Q.~H.}\ \bibnamefont
  {Wang}}, \bibinfo {author} {\bibfnamefont {A.}~\bibnamefont {Bedoya-Pinto}},
  \bibinfo {author} {\bibfnamefont {M.}~\bibnamefont {Blei}}, \bibinfo {author}
  {\bibfnamefont {A.~H.}\ \bibnamefont {Dismukes}}, \bibinfo {author}
  {\bibfnamefont {A.}~\bibnamefont {Hamo}}, \bibinfo {author} {\bibfnamefont
  {S.}~\bibnamefont {Jenkins}}, \bibinfo {author} {\bibfnamefont
  {M.}~\bibnamefont {Koperski}}, \bibinfo {author} {\bibfnamefont
  {Y.}~\bibnamefont {Liu}}, \bibinfo {author} {\bibfnamefont {Q.-C.}\
  \bibnamefont {Sun}}, \bibinfo {author} {\bibfnamefont {E.~J.}\ \bibnamefont
  {Telford}}, \bibinfo {author} {\bibfnamefont {H.~H.}\ \bibnamefont {Kim}},
  \bibinfo {author} {\bibfnamefont {M.}~\bibnamefont {Augustin}}, \bibinfo
  {author} {\bibfnamefont {U.}~\bibnamefont {Vool}}, \bibinfo {author}
  {\bibfnamefont {J.-X.}\ \bibnamefont {Yin}}, \bibinfo {author} {\bibfnamefont
  {L.~H.}\ \bibnamefont {Li}}, \bibinfo {author} {\bibfnamefont
  {A.}~\bibnamefont {Falin}}, \bibinfo {author} {\bibfnamefont {C.~R.}\
  \bibnamefont {Dean}}, \bibinfo {author} {\bibfnamefont {F.}~\bibnamefont
  {Casanova}}, \bibinfo {author} {\bibfnamefont {R.~F.~L.}\ \bibnamefont
  {Evans}}, \bibinfo {author} {\bibfnamefont {M.}~\bibnamefont {Chshiev}},
  \bibinfo {author} {\bibfnamefont {A.}~\bibnamefont {Mishchenko}}, \bibinfo
  {author} {\bibfnamefont {C.}~\bibnamefont {Petrovic}}, \bibinfo {author}
  {\bibfnamefont {R.}~\bibnamefont {He}}, \bibinfo {author} {\bibfnamefont
  {L.}~\bibnamefont {Zhao}}, \bibinfo {author} {\bibfnamefont {A.~W.}\
  \bibnamefont {Tsen}}, \bibinfo {author} {\bibfnamefont {B.~D.}\ \bibnamefont
  {Gerardot}}, \bibinfo {author} {\bibfnamefont {M.}~\bibnamefont
  {Brotons-Gisbert}}, \bibinfo {author} {\bibfnamefont {Z.}~\bibnamefont
  {Guguchia}}, \bibinfo {author} {\bibfnamefont {X.}~\bibnamefont {Roy}},
  \bibinfo {author} {\bibfnamefont {S.}~\bibnamefont {Tongay}}, \bibinfo
  {author} {\bibfnamefont {Z.}~\bibnamefont {Wang}}, \bibinfo {author}
  {\bibfnamefont {M.~Z.}\ \bibnamefont {Hasan}}, \bibinfo {author}
  {\bibfnamefont {J.}~\bibnamefont {Wrachtrup}}, \bibinfo {author}
  {\bibfnamefont {A.}~\bibnamefont {Yacoby}}, \bibinfo {author} {\bibfnamefont
  {A.}~\bibnamefont {Fert}}, \bibinfo {author} {\bibfnamefont {S.}~\bibnamefont
  {Parkin}}, \bibinfo {author} {\bibfnamefont {K.~S.}\ \bibnamefont
  {Novoselov}}, \bibinfo {author} {\bibfnamefont {P.}~\bibnamefont {Dai}},
  \bibinfo {author} {\bibfnamefont {L.}~\bibnamefont {Balicas}},\ and\ \bibinfo
  {author} {\bibfnamefont {E.~J.~G.}\ \bibnamefont {Santos}},\ }\bibfield
  {title} {\enquote {\bibinfo {title} {The magnetic genome of two-dimensional
  van der {W}aals materials},}\ }\href
  {https://doi.org/10.1021/acsnano.1c09150} {\bibfield  {journal} {\bibinfo
  {journal} {ACS Nano}\ }\textbf {\bibinfo {volume} {16}},\ \bibinfo {pages}
  {6960} (\bibinfo {year} {2022})}\BibitemShut {NoStop}%
\bibitem [{\citenamefont {Ziebel}\ \emph {et~al.}(2024)\citenamefont {Ziebel},
  \citenamefont {Feuer}, \citenamefont {Cox}, \citenamefont {Zhu},
  \citenamefont {Dean},\ and\ \citenamefont {Roy}}]{CrSBr-review2024}%
  \BibitemOpen
  \bibfield  {author} {\bibinfo {author} {\bibfnamefont {M.~E.}\ \bibnamefont
  {Ziebel}}, \bibinfo {author} {\bibfnamefont {M.~L.}\ \bibnamefont {Feuer}},
  \bibinfo {author} {\bibfnamefont {J.}~\bibnamefont {Cox}}, \bibinfo {author}
  {\bibfnamefont {X.}~\bibnamefont {Zhu}}, \bibinfo {author} {\bibfnamefont
  {C.~R.}\ \bibnamefont {Dean}},\ and\ \bibinfo {author} {\bibfnamefont
  {X.}~\bibnamefont {Roy}},\ }\bibfield  {title} {\enquote {\bibinfo {title}
  {{CrSBr}: An air-stable, two-dimensional magnetic semiconductor},}\ }\href
  {https://doi.org/10.1021/acs.nanolett.4c00624} {\bibfield  {journal}
  {\bibinfo  {journal} {Nano Letters}\ }\textbf {\bibinfo {volume} {24}},\
  \bibinfo {pages} {4319} (\bibinfo {year} {2024})}\BibitemShut {NoStop}%
\bibitem [{\citenamefont {Tschudin}\ \emph {et~al.}(2024)\citenamefont
  {Tschudin}, \citenamefont {Broadway}, \citenamefont {Siegwolf}, \citenamefont
  {Schrader}, \citenamefont {Telford}, \citenamefont {Gross}, \citenamefont
  {Cox}, \citenamefont {Dubois}, \citenamefont {Chica}, \citenamefont
  {Rama-Eiroa}, \citenamefont {J.~G.~Santos}, \citenamefont {Poggio},
  \citenamefont {Ziebel}, \citenamefont {Dean}, \citenamefont {Roy},\ and\
  \citenamefont {Maletinsky}}]{Maletinski-2024}%
  \BibitemOpen
  \bibfield  {author} {\bibinfo {author} {\bibfnamefont {M.~A.}\ \bibnamefont
  {Tschudin}}, \bibinfo {author} {\bibfnamefont {D.~A.}\ \bibnamefont
  {Broadway}}, \bibinfo {author} {\bibfnamefont {P.}~\bibnamefont {Siegwolf}},
  \bibinfo {author} {\bibfnamefont {C.}~\bibnamefont {Schrader}}, \bibinfo
  {author} {\bibfnamefont {E.~J.}\ \bibnamefont {Telford}}, \bibinfo {author}
  {\bibfnamefont {B.}~\bibnamefont {Gross}}, \bibinfo {author} {\bibfnamefont
  {J.}~\bibnamefont {Cox}}, \bibinfo {author} {\bibfnamefont {A.~E.~E.}\
  \bibnamefont {Dubois}}, \bibinfo {author} {\bibfnamefont {D.~G.}\
  \bibnamefont {Chica}}, \bibinfo {author} {\bibfnamefont {R.}~\bibnamefont
  {Rama-Eiroa}}, \bibinfo {author} {\bibfnamefont {E.}~\bibnamefont
  {J.~G.~Santos}}, \bibinfo {author} {\bibfnamefont {M.}~\bibnamefont
  {Poggio}}, \bibinfo {author} {\bibfnamefont {M.~E.}\ \bibnamefont {Ziebel}},
  \bibinfo {author} {\bibfnamefont {C.~R.}\ \bibnamefont {Dean}}, \bibinfo
  {author} {\bibfnamefont {X.}~\bibnamefont {Roy}},\ and\ \bibinfo {author}
  {\bibfnamefont {P.}~\bibnamefont {Maletinsky}},\ }\bibfield  {title}
  {\enquote {\bibinfo {title} {Imaging nanomagnetism and magnetic phase
  transitions in atomically thin {CrSBr}},}\ }\href
  {https://doi.org/10.1038/s41467-024-49717-9} {\bibfield  {journal} {\bibinfo
  {journal} {Nature Communications}\ }\textbf {\bibinfo {volume} {15}},\
  \bibinfo {pages} {6005} (\bibinfo {year} {2024})}\BibitemShut {NoStop}%
\bibitem [{\citenamefont {Wilson}\ \emph {et~al.}(2021)\citenamefont {Wilson},
  \citenamefont {Lee}, \citenamefont {Cenker}, \citenamefont {Xie},
  \citenamefont {Dismukes}, \citenamefont {Telford}, \citenamefont {Fonseca},
  \citenamefont {Sivakumar}, \citenamefont {Dean}, \citenamefont {Cao},
  \citenamefont {Roy}, \citenamefont {Xu},\ and\ \citenamefont
  {Zhu}}]{Wilson-2021}%
  \BibitemOpen
  \bibfield  {author} {\bibinfo {author} {\bibfnamefont {N.~P.}\ \bibnamefont
  {Wilson}}, \bibinfo {author} {\bibfnamefont {K.}~\bibnamefont {Lee}},
  \bibinfo {author} {\bibfnamefont {J.}~\bibnamefont {Cenker}}, \bibinfo
  {author} {\bibfnamefont {K.}~\bibnamefont {Xie}}, \bibinfo {author}
  {\bibfnamefont {A.~H.}\ \bibnamefont {Dismukes}}, \bibinfo {author}
  {\bibfnamefont {E.~J.}\ \bibnamefont {Telford}}, \bibinfo {author}
  {\bibfnamefont {J.}~\bibnamefont {Fonseca}}, \bibinfo {author} {\bibfnamefont
  {S.}~\bibnamefont {Sivakumar}}, \bibinfo {author} {\bibfnamefont
  {C.}~\bibnamefont {Dean}}, \bibinfo {author} {\bibfnamefont {T.}~\bibnamefont
  {Cao}}, \bibinfo {author} {\bibfnamefont {X.}~\bibnamefont {Roy}}, \bibinfo
  {author} {\bibfnamefont {X.}~\bibnamefont {Xu}},\ and\ \bibinfo {author}
  {\bibfnamefont {X.}~\bibnamefont {Zhu}},\ }\bibfield  {title} {\enquote
  {\bibinfo {title} {Interlayer electronic coupling on demand in a {2D}
  magnetic semiconductor},}\ }\href
  {https://doi.org/10.1038/s41563-021-01070-8} {\bibfield  {journal} {\bibinfo
  {journal} {Nature Materials}\ }\textbf {\bibinfo {volume} {20}},\ \bibinfo
  {pages} {1657} (\bibinfo {year} {2021})}\BibitemShut {NoStop}%
\bibitem [{\citenamefont {Tabataba-Vakili}\ \emph {et~al.}(2024)\citenamefont
  {Tabataba-Vakili}, \citenamefont {Nguyen}, \citenamefont {Rupp},
  \citenamefont {Mosina}, \citenamefont {Papavasileiou}, \citenamefont
  {Watanabe}, \citenamefont {Taniguchi}, \citenamefont {Maletinsky},
  \citenamefont {Glazov}, \citenamefont {Sofer}, \citenamefont {Baimuratov},\
  and\ \citenamefont {H{\"o}gele}}]{Haegele-2024}%
  \BibitemOpen
  \bibfield  {author} {\bibinfo {author} {\bibfnamefont {F.}~\bibnamefont
  {Tabataba-Vakili}}, \bibinfo {author} {\bibfnamefont {H.~P.~G.}\ \bibnamefont
  {Nguyen}}, \bibinfo {author} {\bibfnamefont {A.}~\bibnamefont {Rupp}},
  \bibinfo {author} {\bibfnamefont {K.}~\bibnamefont {Mosina}}, \bibinfo
  {author} {\bibfnamefont {A.}~\bibnamefont {Papavasileiou}}, \bibinfo {author}
  {\bibfnamefont {K.}~\bibnamefont {Watanabe}}, \bibinfo {author}
  {\bibfnamefont {T.}~\bibnamefont {Taniguchi}}, \bibinfo {author}
  {\bibfnamefont {P.}~\bibnamefont {Maletinsky}}, \bibinfo {author}
  {\bibfnamefont {M.~M.}\ \bibnamefont {Glazov}}, \bibinfo {author}
  {\bibfnamefont {Z.}~\bibnamefont {Sofer}}, \bibinfo {author} {\bibfnamefont
  {A.~S.}\ \bibnamefont {Baimuratov}},\ and\ \bibinfo {author} {\bibfnamefont
  {A.}~\bibnamefont {H{\"o}gele}},\ }\bibfield  {title} {\enquote {\bibinfo
  {title} {Doping-control of excitons and magnetism in few-layer {CrSBr}},}\
  }\href {https://doi.org/10.1038/s41467-024-49048-9} {\bibfield  {journal}
  {\bibinfo  {journal} {Nature Communications}\ }\textbf {\bibinfo {volume}
  {15}},\ \bibinfo {pages} {4735} (\bibinfo {year} {2024})}\BibitemShut
  {NoStop}%
\bibitem [{\citenamefont {Bae}\ \emph {et~al.}(2022)\citenamefont {Bae},
  \citenamefont {Wang}, \citenamefont {Scheie}, \citenamefont {Xu},
  \citenamefont {Chica}, \citenamefont {Diederich}, \citenamefont {Cenker},
  \citenamefont {Ziebel}, \citenamefont {Bai}, \citenamefont {Ren},
  \citenamefont {Dean}, \citenamefont {Delor}, \citenamefont {Xu},
  \citenamefont {Roy}, \citenamefont {Kent},\ and\ \citenamefont
  {Zhu}}]{Zhu-magnons-2022}%
  \BibitemOpen
  \bibfield  {author} {\bibinfo {author} {\bibfnamefont {Y.~J.}\ \bibnamefont
  {Bae}}, \bibinfo {author} {\bibfnamefont {J.}~\bibnamefont {Wang}}, \bibinfo
  {author} {\bibfnamefont {A.}~\bibnamefont {Scheie}}, \bibinfo {author}
  {\bibfnamefont {J.}~\bibnamefont {Xu}}, \bibinfo {author} {\bibfnamefont
  {D.~G.}\ \bibnamefont {Chica}}, \bibinfo {author} {\bibfnamefont {G.~M.}\
  \bibnamefont {Diederich}}, \bibinfo {author} {\bibfnamefont {J.}~\bibnamefont
  {Cenker}}, \bibinfo {author} {\bibfnamefont {M.~E.}\ \bibnamefont {Ziebel}},
  \bibinfo {author} {\bibfnamefont {Y.}~\bibnamefont {Bai}}, \bibinfo {author}
  {\bibfnamefont {H.}~\bibnamefont {Ren}}, \bibinfo {author} {\bibfnamefont
  {C.~R.}\ \bibnamefont {Dean}}, \bibinfo {author} {\bibfnamefont
  {M.}~\bibnamefont {Delor}}, \bibinfo {author} {\bibfnamefont
  {X.}~\bibnamefont {Xu}}, \bibinfo {author} {\bibfnamefont {X.}~\bibnamefont
  {Roy}}, \bibinfo {author} {\bibfnamefont {A.~D.}\ \bibnamefont {Kent}},\ and\
  \bibinfo {author} {\bibfnamefont {X.}~\bibnamefont {Zhu}},\ }\bibfield
  {title} {\enquote {\bibinfo {title} {Exciton-coupled coherent magnons in a
  {2D} semiconductor},}\ }\href {https://doi.org/10.1038/s41586-022-05024-1}
  {\bibfield  {journal} {\bibinfo  {journal} {Nature}\ }\textbf {\bibinfo
  {volume} {609}},\ \bibinfo {pages} {282} (\bibinfo {year}
  {2022})}\BibitemShut {NoStop}%
\bibitem [{\citenamefont {Dirnberger}\ \emph {et~al.}(2023)\citenamefont
  {Dirnberger}, \citenamefont {Quan}, \citenamefont {Bushati}, \citenamefont
  {Diederich}, \citenamefont {Florian}, \citenamefont {Klein}, \citenamefont
  {Mosina}, \citenamefont {Sofer}, \citenamefont {Xu}, \citenamefont {Kamra},
  \citenamefont {Garc{\'i}a-Vidal}, \citenamefont {Al{\`u}},\ and\
  \citenamefont {Menon}}]{Dirnberger-2023}%
  \BibitemOpen
  \bibfield  {author} {\bibinfo {author} {\bibfnamefont {F.}~\bibnamefont
  {Dirnberger}}, \bibinfo {author} {\bibfnamefont {J.}~\bibnamefont {Quan}},
  \bibinfo {author} {\bibfnamefont {R.}~\bibnamefont {Bushati}}, \bibinfo
  {author} {\bibfnamefont {G.~M.}\ \bibnamefont {Diederich}}, \bibinfo {author}
  {\bibfnamefont {M.}~\bibnamefont {Florian}}, \bibinfo {author} {\bibfnamefont
  {J.}~\bibnamefont {Klein}}, \bibinfo {author} {\bibfnamefont
  {K.}~\bibnamefont {Mosina}}, \bibinfo {author} {\bibfnamefont
  {Z.}~\bibnamefont {Sofer}}, \bibinfo {author} {\bibfnamefont
  {X.}~\bibnamefont {Xu}}, \bibinfo {author} {\bibfnamefont {A.}~\bibnamefont
  {Kamra}}, \bibinfo {author} {\bibfnamefont {F.~J.}\ \bibnamefont
  {Garc{\'i}a-Vidal}}, \bibinfo {author} {\bibfnamefont {A.}~\bibnamefont
  {Al{\`u}}},\ and\ \bibinfo {author} {\bibfnamefont {V.~M.}\ \bibnamefont
  {Menon}},\ }\bibfield  {title} {\enquote {\bibinfo {title} {Magneto-optics in
  a van der {W}aals magnet tuned by self-hybridized polaritons},}\ }\href
  {https://doi.org/10.1038/s41586-023-06275-2} {\bibfield  {journal} {\bibinfo
  {journal} {Nature}\ }\textbf {\bibinfo {volume} {620}},\ \bibinfo {pages}
  {533} (\bibinfo {year} {2023})}\BibitemShut {NoStop}%
\bibitem [{\citenamefont {Wang}\ \emph {et~al.}(2023)\citenamefont {Wang},
  \citenamefont {Zhang}, \citenamefont {Yang}, \citenamefont {Lin},
  \citenamefont {Chen}, \citenamefont {Yang}, \citenamefont {Gong},
  \citenamefont {Chen}, \citenamefont {Ye},\ and\ \citenamefont
  {Liu}}]{Wang2023}%
  \BibitemOpen
  \bibfield  {author} {\bibinfo {author} {\bibfnamefont {T.}~\bibnamefont
  {Wang}}, \bibinfo {author} {\bibfnamefont {D.}~\bibnamefont {Zhang}},
  \bibinfo {author} {\bibfnamefont {S.}~\bibnamefont {Yang}}, \bibinfo {author}
  {\bibfnamefont {Z.}~\bibnamefont {Lin}}, \bibinfo {author} {\bibfnamefont
  {Q.}~\bibnamefont {Chen}}, \bibinfo {author} {\bibfnamefont {J.}~\bibnamefont
  {Yang}}, \bibinfo {author} {\bibfnamefont {Q.}~\bibnamefont {Gong}}, \bibinfo
  {author} {\bibfnamefont {Z.}~\bibnamefont {Chen}}, \bibinfo {author}
  {\bibfnamefont {Y.}~\bibnamefont {Ye}},\ and\ \bibinfo {author}
  {\bibfnamefont {W.}~\bibnamefont {Liu}},\ }\bibfield  {title} {\enquote
  {\bibinfo {title} {Magnetically-dressed {CrSBr} exciton-polaritons in
  ultrastrong coupling regime},}\ }\href
  {https://doi.org/10.1038/s41467-023-41688-7} {\bibfield  {journal} {\bibinfo
  {journal} {Nature Communications}\ }\textbf {\bibinfo {volume} {14}},\
  \bibinfo {pages} {5966} (\bibinfo {year} {2023})}\BibitemShut {NoStop}%
\bibitem [{\citenamefont {Ivchenko}(2005)}]{Ivchenko-book}%
  \BibitemOpen
  \bibfield  {author} {\bibinfo {author} {\bibfnamefont {E.~L.}\ \bibnamefont
  {Ivchenko}},\ }\href@noop {} {\emph {\bibinfo {title} {Optical spectroscopy
  of semiconductor nanostructures}}}\ (\bibinfo  {publisher} {Alpha Science,
  Harrow UK},\ \bibinfo {year} {2005})\BibitemShut {NoStop}%
\bibitem [{\citenamefont {Klein}\ \emph
  {et~al.}(2023{\natexlab{a}})\citenamefont {Klein}, \citenamefont {Pingault},
  \citenamefont {Florian}, \citenamefont {Heißenb{\"u}ttel}, \citenamefont
  {Steinhoff}, \citenamefont {Song}, \citenamefont {Torres}, \citenamefont
  {Dirnberger}, \citenamefont {Curtis}, \citenamefont {Weile}, \citenamefont
  {Penn}, \citenamefont {Deilmann}, \citenamefont {Dana}, \citenamefont
  {Bushati}, \citenamefont {Quan}, \citenamefont {Luxa}, \citenamefont {Sofer},
  \citenamefont {Alù}, \citenamefont {Menon}, \citenamefont {Wurstbauer},
  \citenamefont {Rohlfing}, \citenamefont {Narang}, \citenamefont {Lončar},\
  and\ \citenamefont {Ross}}]{Klein-2023}%
  \BibitemOpen
  \bibfield  {author} {\bibinfo {author} {\bibfnamefont {J.}~\bibnamefont
  {Klein}}, \bibinfo {author} {\bibfnamefont {B.}~\bibnamefont {Pingault}},
  \bibinfo {author} {\bibfnamefont {M.}~\bibnamefont {Florian}}, \bibinfo
  {author} {\bibfnamefont {M.-C.}\ \bibnamefont {Heißenb{\"u}ttel}}, \bibinfo
  {author} {\bibfnamefont {A.}~\bibnamefont {Steinhoff}}, \bibinfo {author}
  {\bibfnamefont {Z.}~\bibnamefont {Song}}, \bibinfo {author} {\bibfnamefont
  {K.}~\bibnamefont {Torres}}, \bibinfo {author} {\bibfnamefont
  {F.}~\bibnamefont {Dirnberger}}, \bibinfo {author} {\bibfnamefont {J.~B.}\
  \bibnamefont {Curtis}}, \bibinfo {author} {\bibfnamefont {M.}~\bibnamefont
  {Weile}}, \bibinfo {author} {\bibfnamefont {A.}~\bibnamefont {Penn}},
  \bibinfo {author} {\bibfnamefont {T.}~\bibnamefont {Deilmann}}, \bibinfo
  {author} {\bibfnamefont {R.}~\bibnamefont {Dana}}, \bibinfo {author}
  {\bibfnamefont {R.}~\bibnamefont {Bushati}}, \bibinfo {author} {\bibfnamefont
  {J.}~\bibnamefont {Quan}}, \bibinfo {author} {\bibfnamefont {J.}~\bibnamefont
  {Luxa}}, \bibinfo {author} {\bibfnamefont {Z.}~\bibnamefont {Sofer}},
  \bibinfo {author} {\bibfnamefont {A.}~\bibnamefont {Alù}}, \bibinfo {author}
  {\bibfnamefont {V.~M.}\ \bibnamefont {Menon}}, \bibinfo {author}
  {\bibfnamefont {U.}~\bibnamefont {Wurstbauer}}, \bibinfo {author}
  {\bibfnamefont {M.}~\bibnamefont {Rohlfing}}, \bibinfo {author}
  {\bibfnamefont {P.}~\bibnamefont {Narang}}, \bibinfo {author} {\bibfnamefont
  {M.}~\bibnamefont {Lončar}},\ and\ \bibinfo {author} {\bibfnamefont {F.~M.}\
  \bibnamefont {Ross}},\ }\bibfield  {title} {\enquote {\bibinfo {title} {The
  bulk van der {W}aals layered magnet {CrSBr} is a quasi-{1D} material},}\
  }\href {https://doi.org/10.1021/acsnano.2c07316} {\bibfield  {journal}
  {\bibinfo  {journal} {ACS Nano}\ }\textbf {\bibinfo {volume} {17}},\ \bibinfo
  {pages} {5316} (\bibinfo {year} {2023}{\natexlab{a}})}\BibitemShut {NoStop}%
\bibitem [{\citenamefont {Telford}\ \emph {et~al.}(2022)\citenamefont
  {Telford}, \citenamefont {Dismukes}, \citenamefont {Dudley}, \citenamefont
  {Wiscons}, \citenamefont {Lee}, \citenamefont {Chica}, \citenamefont
  {Ziebel}, \citenamefont {Han}, \citenamefont {Yu}, \citenamefont {Shabani},
  \citenamefont {Scheie}, \citenamefont {Watanabe}, \citenamefont {Taniguchi},
  \citenamefont {Xiao}, \citenamefont {Zhu}, \citenamefont {Pasupathy},
  \citenamefont {Nuckolls}, \citenamefont {Zhu}, \citenamefont {Dean},\ and\
  \citenamefont {Roy}}]{Telford-2022}%
  \BibitemOpen
  \bibfield  {author} {\bibinfo {author} {\bibfnamefont {E.~J.}\ \bibnamefont
  {Telford}}, \bibinfo {author} {\bibfnamefont {A.~H.}\ \bibnamefont
  {Dismukes}}, \bibinfo {author} {\bibfnamefont {R.~L.}\ \bibnamefont
  {Dudley}}, \bibinfo {author} {\bibfnamefont {R.~A.}\ \bibnamefont {Wiscons}},
  \bibinfo {author} {\bibfnamefont {K.}~\bibnamefont {Lee}}, \bibinfo {author}
  {\bibfnamefont {D.~G.}\ \bibnamefont {Chica}}, \bibinfo {author}
  {\bibfnamefont {M.~E.}\ \bibnamefont {Ziebel}}, \bibinfo {author}
  {\bibfnamefont {M.-G.}\ \bibnamefont {Han}}, \bibinfo {author} {\bibfnamefont
  {J.}~\bibnamefont {Yu}}, \bibinfo {author} {\bibfnamefont {S.}~\bibnamefont
  {Shabani}}, \bibinfo {author} {\bibfnamefont {A.}~\bibnamefont {Scheie}},
  \bibinfo {author} {\bibfnamefont {K.}~\bibnamefont {Watanabe}}, \bibinfo
  {author} {\bibfnamefont {T.}~\bibnamefont {Taniguchi}}, \bibinfo {author}
  {\bibfnamefont {D.}~\bibnamefont {Xiao}}, \bibinfo {author} {\bibfnamefont
  {Y.}~\bibnamefont {Zhu}}, \bibinfo {author} {\bibfnamefont {A.~N.}\
  \bibnamefont {Pasupathy}}, \bibinfo {author} {\bibfnamefont {C.}~\bibnamefont
  {Nuckolls}}, \bibinfo {author} {\bibfnamefont {X.}~\bibnamefont {Zhu}},
  \bibinfo {author} {\bibfnamefont {C.~R.}\ \bibnamefont {Dean}},\ and\
  \bibinfo {author} {\bibfnamefont {X.}~\bibnamefont {Roy}},\ }\bibfield
  {title} {\enquote {\bibinfo {title} {Coupling between magnetic order and
  charge transport in a two-dimensional magnetic semiconductor},}\ }\href
  {https://doi.org/10.1038/s41563-022-01245-x} {\bibfield  {journal} {\bibinfo
  {journal} {Nature Materials}\ }\textbf {\bibinfo {volume} {21}},\ \bibinfo
  {pages} {754} (\bibinfo {year} {2022})}\BibitemShut {NoStop}%
\bibitem [{\citenamefont {Meineke}\ \emph {et~al.}(2024)\citenamefont
  {Meineke}, \citenamefont {Schlosser}, \citenamefont {Zizlsperger},
  \citenamefont {Liebich}, \citenamefont {Nilforoushan}, \citenamefont
  {Mosina}, \citenamefont {Terres}, \citenamefont {Chernikov}, \citenamefont
  {Sofer}, \citenamefont {Huber}, \citenamefont {Florian}, \citenamefont
  {Kira}, \citenamefont {Dirnberger},\ and\ \citenamefont
  {Huber}}]{Meineke-2024}%
  \BibitemOpen
  \bibfield  {author} {\bibinfo {author} {\bibfnamefont {C.}~\bibnamefont
  {Meineke}}, \bibinfo {author} {\bibfnamefont {J.}~\bibnamefont {Schlosser}},
  \bibinfo {author} {\bibfnamefont {M.}~\bibnamefont {Zizlsperger}}, \bibinfo
  {author} {\bibfnamefont {M.}~\bibnamefont {Liebich}}, \bibinfo {author}
  {\bibfnamefont {N.}~\bibnamefont {Nilforoushan}}, \bibinfo {author}
  {\bibfnamefont {K.}~\bibnamefont {Mosina}}, \bibinfo {author} {\bibfnamefont
  {S.}~\bibnamefont {Terres}}, \bibinfo {author} {\bibfnamefont
  {A.}~\bibnamefont {Chernikov}}, \bibinfo {author} {\bibfnamefont
  {Z.}~\bibnamefont {Sofer}}, \bibinfo {author} {\bibfnamefont {M.~A.}\
  \bibnamefont {Huber}}, \bibinfo {author} {\bibfnamefont {M.}~\bibnamefont
  {Florian}}, \bibinfo {author} {\bibfnamefont {M.}~\bibnamefont {Kira}},
  \bibinfo {author} {\bibfnamefont {F.}~\bibnamefont {Dirnberger}},\ and\
  \bibinfo {author} {\bibfnamefont {R.}~\bibnamefont {Huber}},\ }\bibfield
  {title} {\enquote {\bibinfo {title} {Ultrafast exciton dynamics in the
  atomically thin van der {W}aals magnet {CrSBr}},}\ }\href
  {https://doi.org/10.1021/acs.nanolett.3c05010} {\bibfield  {journal}
  {\bibinfo  {journal} {Nano Letters}\ }\textbf {\bibinfo {volume} {24}},\
  \bibinfo {pages} {4101} (\bibinfo {year} {2024})}\BibitemShut {NoStop}%
\bibitem [{\citenamefont {Lin}\ \emph {et~al.}(2024{\natexlab{a}})\citenamefont
  {Lin}, \citenamefont {Sun}, \citenamefont {Dirnberger}, \citenamefont {Li},
  \citenamefont {Qu}, \citenamefont {Wen}, \citenamefont {Sofer}, \citenamefont
  {S{\"o}ll}, \citenamefont {Winnerl}, \citenamefont {Helm}, \citenamefont
  {Zhou}, \citenamefont {Dan},\ and\ \citenamefont {Prucnal}}]{Lin-2024}%
  \BibitemOpen
  \bibfield  {author} {\bibinfo {author} {\bibfnamefont {K.}~\bibnamefont
  {Lin}}, \bibinfo {author} {\bibfnamefont {X.}~\bibnamefont {Sun}}, \bibinfo
  {author} {\bibfnamefont {F.}~\bibnamefont {Dirnberger}}, \bibinfo {author}
  {\bibfnamefont {Y.}~\bibnamefont {Li}}, \bibinfo {author} {\bibfnamefont
  {J.}~\bibnamefont {Qu}}, \bibinfo {author} {\bibfnamefont {P.}~\bibnamefont
  {Wen}}, \bibinfo {author} {\bibfnamefont {Z.}~\bibnamefont {Sofer}}, \bibinfo
  {author} {\bibfnamefont {A.}~\bibnamefont {S{\"o}ll}}, \bibinfo {author}
  {\bibfnamefont {S.}~\bibnamefont {Winnerl}}, \bibinfo {author} {\bibfnamefont
  {M.}~\bibnamefont {Helm}}, \bibinfo {author} {\bibfnamefont {S.}~\bibnamefont
  {Zhou}}, \bibinfo {author} {\bibfnamefont {Y.}~\bibnamefont {Dan}},\ and\
  \bibinfo {author} {\bibfnamefont {S.}~\bibnamefont {Prucnal}},\ }\bibfield
  {title} {\enquote {\bibinfo {title} {Strong exciton--phonon coupling as a
  fingerprint of magnetic ordering in van der waals layered {CrSBr}},}\ }\href
  {https://doi.org/10.1021/acsnano.3c07236} {\bibfield  {journal} {\bibinfo
  {journal} {ACS Nano}\ }\textbf {\bibinfo {volume} {18}},\ \bibinfo {pages}
  {2898} (\bibinfo {year} {2024}{\natexlab{a}})}\BibitemShut {NoStop}%
\bibitem [{\citenamefont {Lin}\ \emph {et~al.}(2024{\natexlab{b}})\citenamefont
  {Lin}, \citenamefont {Li}, \citenamefont {Ghorbani-Asl}, \citenamefont
  {Sofer}, \citenamefont {Winnerl}, \citenamefont {Erbe}, \citenamefont
  {Krasheninnikov}, \citenamefont {Helm}, \citenamefont {Zhou}, \citenamefont
  {Dan} \emph {et~al.}}]{Lin-2024b}%
  \BibitemOpen
  \bibfield  {author} {\bibinfo {author} {\bibfnamefont {K.}~\bibnamefont
  {Lin}}, \bibinfo {author} {\bibfnamefont {Y.}~\bibnamefont {Li}}, \bibinfo
  {author} {\bibfnamefont {M.}~\bibnamefont {Ghorbani-Asl}}, \bibinfo {author}
  {\bibfnamefont {Z.}~\bibnamefont {Sofer}}, \bibinfo {author} {\bibfnamefont
  {S.}~\bibnamefont {Winnerl}}, \bibinfo {author} {\bibfnamefont
  {A.}~\bibnamefont {Erbe}}, \bibinfo {author} {\bibfnamefont {A.~V.}\
  \bibnamefont {Krasheninnikov}}, \bibinfo {author} {\bibfnamefont
  {M.}~\bibnamefont {Helm}}, \bibinfo {author} {\bibfnamefont {S.}~\bibnamefont
  {Zhou}}, \bibinfo {author} {\bibfnamefont {Y.}~\bibnamefont {Dan}}, \emph
  {et~al.},\ }\bibfield  {title} {\enquote {\bibinfo {title} {Probing the band
  splitting near the {$\Gamma$} point in the van der waals magnetic
  semiconductor {CrSBr}},}\ }\href
  {https://doi.org/10.1021/acs.jpclett.4c00968} {\bibfield  {journal} {\bibinfo
   {journal} {The Journal of Physical Chemistry Letters}\ }\textbf {\bibinfo
  {volume} {15}},\ \bibinfo {pages} {6010} (\bibinfo {year}
  {2024}{\natexlab{b}})}\BibitemShut {NoStop}%
\bibitem [{\citenamefont {Krelle}\ \emph {et~al.}(2025)\citenamefont {Krelle},
  \citenamefont {Tan}, \citenamefont {Markina}, \citenamefont {Mondal},
  \citenamefont {Mosina}, \citenamefont {Hagmann}, \citenamefont {von
  Klitzing}, \citenamefont {Watanabe}, \citenamefont {Taniguchi}, \citenamefont
  {Sofer},\ and\ \citenamefont {Urbaszek}}]{Urbaszek-2025}%
  \BibitemOpen
  \bibfield  {author} {\bibinfo {author} {\bibfnamefont {L.}~\bibnamefont
  {Krelle}}, \bibinfo {author} {\bibfnamefont {R.}~\bibnamefont {Tan}},
  \bibinfo {author} {\bibfnamefont {D.}~\bibnamefont {Markina}}, \bibinfo
  {author} {\bibfnamefont {P.}~\bibnamefont {Mondal}}, \bibinfo {author}
  {\bibfnamefont {K.}~\bibnamefont {Mosina}}, \bibinfo {author} {\bibfnamefont
  {K.}~\bibnamefont {Hagmann}}, \bibinfo {author} {\bibfnamefont
  {R.}~\bibnamefont {von Klitzing}}, \bibinfo {author} {\bibfnamefont
  {K.}~\bibnamefont {Watanabe}}, \bibinfo {author} {\bibfnamefont
  {T.}~\bibnamefont {Taniguchi}}, \bibinfo {author} {\bibfnamefont
  {Z.}~\bibnamefont {Sofer}},\ and\ \bibinfo {author} {\bibfnamefont
  {B.}~\bibnamefont {Urbaszek}},\ }\bibfield  {title} {\enquote {\bibinfo
  {title} {Magnetic correlation spectroscopy in {CrSBr}},}\ }\href
  {https://doi.org/10.1021/acsnano.5c05470} {\bibfield  {journal} {\bibinfo
  {journal} {ACS Nano}\ }\textbf {\bibinfo {volume} {19}},\ \bibinfo {pages}
  {33156} (\bibinfo {year} {2025})}\BibitemShut {NoStop}%
\bibitem [{\citenamefont {Permogorov}(1982)}]{Permogorov-Excitons}%
  \BibitemOpen
  \bibfield  {author} {\bibinfo {author} {\bibfnamefont {S.}~\bibnamefont
  {Permogorov}},\ }\bibfield  {title} {\enquote {\bibinfo {title} {Chapter
  5},}\ }in\ \href@noop {} {\emph {\bibinfo {booktitle} {Excitons}}},\ \bibinfo
  {series and number} {Modern Problems in Solid State Physics Series},\
  \bibinfo {editor} {edited by\ \bibinfo {editor} {\bibfnamefont {E.~I.}\
  \bibnamefont {Rashba}}\ and\ \bibinfo {editor} {\bibfnamefont {M.~D.}\
  \bibnamefont {Sturge}}}\ (\bibinfo  {publisher} {North-Holland Publihing
  Company},\ \bibinfo {year} {1982})\BibitemShut {NoStop}%
\bibitem [{\citenamefont {Telford}\ \emph {et~al.}(2020)\citenamefont
  {Telford}, \citenamefont {Dismukes}, \citenamefont {Lee}, \citenamefont
  {Cheng}, \citenamefont {Wieteska}, \citenamefont {Bartholomew}, \citenamefont
  {Chen}, \citenamefont {Xu}, \citenamefont {Pasupathy}, \citenamefont {Zhu},
  \citenamefont {Dean},\ and\ \citenamefont {Roy}}]{Telford-2020}%
  \BibitemOpen
  \bibfield  {author} {\bibinfo {author} {\bibfnamefont {E.~J.}\ \bibnamefont
  {Telford}}, \bibinfo {author} {\bibfnamefont {A.~H.}\ \bibnamefont
  {Dismukes}}, \bibinfo {author} {\bibfnamefont {K.}~\bibnamefont {Lee}},
  \bibinfo {author} {\bibfnamefont {M.}~\bibnamefont {Cheng}}, \bibinfo
  {author} {\bibfnamefont {A.}~\bibnamefont {Wieteska}}, \bibinfo {author}
  {\bibfnamefont {A.~K.}\ \bibnamefont {Bartholomew}}, \bibinfo {author}
  {\bibfnamefont {Y.-S.}\ \bibnamefont {Chen}}, \bibinfo {author}
  {\bibfnamefont {X.}~\bibnamefont {Xu}}, \bibinfo {author} {\bibfnamefont
  {A.~N.}\ \bibnamefont {Pasupathy}}, \bibinfo {author} {\bibfnamefont
  {X.}~\bibnamefont {Zhu}}, \bibinfo {author} {\bibfnamefont {C.~R.}\
  \bibnamefont {Dean}},\ and\ \bibinfo {author} {\bibfnamefont
  {X.}~\bibnamefont {Roy}},\ }\bibfield  {title} {\enquote {\bibinfo {title}
  {Layered antiferromagnetism induces large negative magnetoresistance in the
  van der {W}aals semiconductor {CrSBr}},}\ }\href
  {https://doi.org/https://doi.org/10.1002/adma.202003240} {\bibfield
  {journal} {\bibinfo  {journal} {Advanced Materials}\ }\textbf {\bibinfo
  {volume} {32}},\ \bibinfo {pages} {2003240} (\bibinfo {year}
  {2020})}\BibitemShut {NoStop}%
\bibitem [{\citenamefont {Klein}\ \emph
  {et~al.}(2023{\natexlab{b}})\citenamefont {Klein}, \citenamefont {Song},
  \citenamefont {Pingault}, \citenamefont {Dirnberger}, \citenamefont {Chi},
  \citenamefont {Curtis}, \citenamefont {Dana}, \citenamefont {Bushati},
  \citenamefont {Quan}, \citenamefont {Dekanovsky}, \citenamefont {Sofer},
  \citenamefont {Alù}, \citenamefont {Menon}, \citenamefont {Moodera},
  \citenamefont {Lončar}, \citenamefont {Narang},\ and\ \citenamefont
  {Ross}}]{Klein-2023s}%
  \BibitemOpen
  \bibfield  {author} {\bibinfo {author} {\bibfnamefont {J.}~\bibnamefont
  {Klein}}, \bibinfo {author} {\bibfnamefont {Z.}~\bibnamefont {Song}},
  \bibinfo {author} {\bibfnamefont {B.}~\bibnamefont {Pingault}}, \bibinfo
  {author} {\bibfnamefont {F.}~\bibnamefont {Dirnberger}}, \bibinfo {author}
  {\bibfnamefont {H.}~\bibnamefont {Chi}}, \bibinfo {author} {\bibfnamefont
  {J.~B.}\ \bibnamefont {Curtis}}, \bibinfo {author} {\bibfnamefont
  {R.}~\bibnamefont {Dana}}, \bibinfo {author} {\bibfnamefont {R.}~\bibnamefont
  {Bushati}}, \bibinfo {author} {\bibfnamefont {J.}~\bibnamefont {Quan}},
  \bibinfo {author} {\bibfnamefont {L.}~\bibnamefont {Dekanovsky}}, \bibinfo
  {author} {\bibfnamefont {Z.}~\bibnamefont {Sofer}}, \bibinfo {author}
  {\bibfnamefont {A.}~\bibnamefont {Alù}}, \bibinfo {author} {\bibfnamefont
  {V.~M.}\ \bibnamefont {Menon}}, \bibinfo {author} {\bibfnamefont {J.~S.}\
  \bibnamefont {Moodera}}, \bibinfo {author} {\bibfnamefont {M.}~\bibnamefont
  {Lončar}}, \bibinfo {author} {\bibfnamefont {P.}~\bibnamefont {Narang}},\
  and\ \bibinfo {author} {\bibfnamefont {F.~M.}\ \bibnamefont {Ross}},\
  }\bibfield  {title} {\enquote {\bibinfo {title} {Sensing the local magnetic
  environment through optically active defects in a layered magnetic
  semiconductor},}\ }\href {https://doi.org/10.1021/acsnano.2c07655} {\bibfield
   {journal} {\bibinfo  {journal} {ACS Nano}\ }\textbf {\bibinfo {volume}
  {17}},\ \bibinfo {pages} {288} (\bibinfo {year}
  {2023}{\natexlab{b}})}\BibitemShut {NoStop}%
\bibitem [{\citenamefont {Léonard}\ \emph {et~al.}(2014)\citenamefont
  {Léonard}, \citenamefont {Dumas}, \citenamefont {Caussé}, \citenamefont
  {Maillot}, \citenamefont {Giannakopoulou}, \citenamefont {Barre},\ and\
  \citenamefont {Uhring}}]{IRF-2016}%
  \BibitemOpen
  \bibfield  {author} {\bibinfo {author} {\bibfnamefont {J.}~\bibnamefont
  {Léonard}}, \bibinfo {author} {\bibfnamefont {N.}~\bibnamefont {Dumas}},
  \bibinfo {author} {\bibfnamefont {J.-P.}\ \bibnamefont {Caussé}}, \bibinfo
  {author} {\bibfnamefont {S.}~\bibnamefont {Maillot}}, \bibinfo {author}
  {\bibfnamefont {N.}~\bibnamefont {Giannakopoulou}}, \bibinfo {author}
  {\bibfnamefont {S.}~\bibnamefont {Barre}},\ and\ \bibinfo {author}
  {\bibfnamefont {W.}~\bibnamefont {Uhring}},\ }\bibfield  {title} {\enquote
  {\bibinfo {title} {High-throughput time-correlated single photon counting},}\
  }\href {https://doi.org/10.1039/C4LC00780H} {\bibfield  {journal} {\bibinfo
  {journal} {Lab Chip}\ }\textbf {\bibinfo {volume} {14}},\ \bibinfo {pages}
  {4338} (\bibinfo {year} {2014})}\BibitemShut {NoStop}%
\bibitem [{\citenamefont {Semina}\ \emph {et~al.}(2025)\citenamefont {Semina},
  \citenamefont {Tabataba-Vakili}, \citenamefont {Rupp}, \citenamefont
  {Baimuratov}, \citenamefont {H\"ogele},\ and\ \citenamefont
  {Glazov}}]{Semina2025}%
  \BibitemOpen
  \bibfield  {author} {\bibinfo {author} {\bibfnamefont {M.~A.}\ \bibnamefont
  {Semina}}, \bibinfo {author} {\bibfnamefont {F.}~\bibnamefont
  {Tabataba-Vakili}}, \bibinfo {author} {\bibfnamefont {A.}~\bibnamefont
  {Rupp}}, \bibinfo {author} {\bibfnamefont {A.~S.}\ \bibnamefont
  {Baimuratov}}, \bibinfo {author} {\bibfnamefont {A.}~\bibnamefont
  {H\"ogele}},\ and\ \bibinfo {author} {\bibfnamefont {M.~M.}\ \bibnamefont
  {Glazov}},\ }\bibfield  {title} {\enquote {\bibinfo {title} {Excitons and
  trions in {CrSBr} bilayers},}\ }\href
  {https://doi.org/10.1103/PhysRevB.111.205301} {\bibfield  {journal} {\bibinfo
   {journal} {Phys. Rev. B}\ }\textbf {\bibinfo {volume} {111}},\ \bibinfo
  {pages} {205301} (\bibinfo {year} {2025})}\BibitemShut {NoStop}%
\bibitem [{\citenamefont {Fang}\ \emph {et~al.}(2019)\citenamefont {Fang},
  \citenamefont {Han}, \citenamefont {Robert}, \citenamefont {Semina},
  \citenamefont {Lagarde}, \citenamefont {Courtade}, \citenamefont {Taniguchi},
  \citenamefont {Watanabe}, \citenamefont {Amand}, \citenamefont {Urbaszek},
  \citenamefont {Glazov},\ and\ \citenamefont {Marie}}]{Fang-2019}%
  \BibitemOpen
  \bibfield  {author} {\bibinfo {author} {\bibfnamefont {H.~H.}\ \bibnamefont
  {Fang}}, \bibinfo {author} {\bibfnamefont {B.}~\bibnamefont {Han}}, \bibinfo
  {author} {\bibfnamefont {C.}~\bibnamefont {Robert}}, \bibinfo {author}
  {\bibfnamefont {M.~A.}\ \bibnamefont {Semina}}, \bibinfo {author}
  {\bibfnamefont {D.}~\bibnamefont {Lagarde}}, \bibinfo {author} {\bibfnamefont
  {E.}~\bibnamefont {Courtade}}, \bibinfo {author} {\bibfnamefont
  {T.}~\bibnamefont {Taniguchi}}, \bibinfo {author} {\bibfnamefont
  {K.}~\bibnamefont {Watanabe}}, \bibinfo {author} {\bibfnamefont
  {T.}~\bibnamefont {Amand}}, \bibinfo {author} {\bibfnamefont
  {B.}~\bibnamefont {Urbaszek}}, \bibinfo {author} {\bibfnamefont {M.~M.}\
  \bibnamefont {Glazov}},\ and\ \bibinfo {author} {\bibfnamefont
  {X.}~\bibnamefont {Marie}},\ }\bibfield  {title} {\enquote {\bibinfo {title}
  {Control of the exciton radiative lifetime in van der {W}aals
  heterostructures},}\ }\href {https://doi.org/10.1103/PhysRevLett.123.067401}
  {\bibfield  {journal} {\bibinfo  {journal} {Phys. Rev. Lett.}\ }\textbf
  {\bibinfo {volume} {123}},\ \bibinfo {pages} {067401} (\bibinfo {year}
  {2019})}\BibitemShut {NoStop}%
\bibitem [{\citenamefont {Korn}\ \emph {et~al.}(2011)\citenamefont {Korn},
  \citenamefont {Heydrich}, \citenamefont {Hirmer}, \citenamefont
  {Schmutzler},\ and\ \citenamefont {Schüller}}]{Korn-2011}%
  \BibitemOpen
  \bibfield  {author} {\bibinfo {author} {\bibfnamefont {T.}~\bibnamefont
  {Korn}}, \bibinfo {author} {\bibfnamefont {S.}~\bibnamefont {Heydrich}},
  \bibinfo {author} {\bibfnamefont {M.}~\bibnamefont {Hirmer}}, \bibinfo
  {author} {\bibfnamefont {J.}~\bibnamefont {Schmutzler}},\ and\ \bibinfo
  {author} {\bibfnamefont {C.}~\bibnamefont {Schüller}},\ }\bibfield  {title}
  {\enquote {\bibinfo {title} {Low-temperature photocarrier dynamics in
  monolayer {MoS\texorpdfstring{\textsubscript{2}}{2}}},}\ }\href
  {https://doi.org/10.1063/1.3636402} {\bibfield  {journal} {\bibinfo
  {journal} {Applied Physics Letters}\ }\textbf {\bibinfo {volume} {99}},\
  \bibinfo {pages} {102109} (\bibinfo {year} {2011})}\BibitemShut {NoStop}%
\bibitem [{\citenamefont {Robert}\ \emph {et~al.}(2016)\citenamefont {Robert},
  \citenamefont {Lagarde}, \citenamefont {Cadiz}, \citenamefont {Wang},
  \citenamefont {Lassagne}, \citenamefont {Amand}, \citenamefont {Balocchi},
  \citenamefont {Renucci}, \citenamefont {Tongay}, \citenamefont {Urbaszek},\
  and\ \citenamefont {Marie}}]{Robert-2016}%
  \BibitemOpen
  \bibfield  {author} {\bibinfo {author} {\bibfnamefont {C.}~\bibnamefont
  {Robert}}, \bibinfo {author} {\bibfnamefont {D.}~\bibnamefont {Lagarde}},
  \bibinfo {author} {\bibfnamefont {F.}~\bibnamefont {Cadiz}}, \bibinfo
  {author} {\bibfnamefont {G.}~\bibnamefont {Wang}}, \bibinfo {author}
  {\bibfnamefont {B.}~\bibnamefont {Lassagne}}, \bibinfo {author}
  {\bibfnamefont {T.}~\bibnamefont {Amand}}, \bibinfo {author} {\bibfnamefont
  {A.}~\bibnamefont {Balocchi}}, \bibinfo {author} {\bibfnamefont
  {P.}~\bibnamefont {Renucci}}, \bibinfo {author} {\bibfnamefont
  {S.}~\bibnamefont {Tongay}}, \bibinfo {author} {\bibfnamefont
  {B.}~\bibnamefont {Urbaszek}},\ and\ \bibinfo {author} {\bibfnamefont
  {X.}~\bibnamefont {Marie}},\ }\bibfield  {title} {\enquote {\bibinfo {title}
  {Exciton radiative lifetime in transition metal dichalcogenide monolayers},}\
  }\href {https://doi.org/10.1103/PhysRevB.93.205423} {\bibfield  {journal}
  {\bibinfo  {journal} {Phys. Rev. B}\ }\textbf {\bibinfo {volume} {93}},\
  \bibinfo {pages} {205423} (\bibinfo {year} {2016})}\BibitemShut {NoStop}%
\bibitem [{\citenamefont {Kossut}\ and\ \citenamefont
  {Gaj}(2010)}]{Kossut-book}%
  \BibitemOpen
  \bibinfo {editor} {\bibfnamefont {J.}~\bibnamefont {Kossut}}\ and\ \bibinfo
  {editor} {\bibfnamefont {J.~A.}\ \bibnamefont {Gaj}},\ eds.,\ \href@noop {}
  {\emph {\bibinfo {title} {Introduction to the Physics of Diluted Magnetic
  Semiconductors}}}\ (\bibinfo  {publisher} {Springer, Berlin},\ \bibinfo
  {year} {2010})\BibitemShut {NoStop}%
\bibitem [{\citenamefont {Citrin}(1992)}]{Citrin1992}%
  \BibitemOpen
  \bibfield  {author} {\bibinfo {author} {\bibfnamefont {D.~S.}\ \bibnamefont
  {Citrin}},\ }\bibfield  {title} {\enquote {\bibinfo {title} {Long intrinsic
  radiative lifetimes of excitons in quantum wires},}\ }\href
  {https://doi.org/10.1103/PhysRevLett.69.3393} {\bibfield  {journal} {\bibinfo
   {journal} {Phys. Rev. Lett.}\ }\textbf {\bibinfo {volume} {69}},\ \bibinfo
  {pages} {3393} (\bibinfo {year} {1992})}\BibitemShut {NoStop}%
\bibitem [{\citenamefont {Abrahams}\ \emph {et~al.}(1979)\citenamefont
  {Abrahams}, \citenamefont {Anderson}, \citenamefont {Licciardello},\ and\
  \citenamefont {Ramakrishnan}}]{Abrahams-1979}%
  \BibitemOpen
  \bibfield  {author} {\bibinfo {author} {\bibfnamefont {E.}~\bibnamefont
  {Abrahams}}, \bibinfo {author} {\bibfnamefont {P.~W.}\ \bibnamefont
  {Anderson}}, \bibinfo {author} {\bibfnamefont {D.~C.}\ \bibnamefont
  {Licciardello}},\ and\ \bibinfo {author} {\bibfnamefont {T.~V.}\ \bibnamefont
  {Ramakrishnan}},\ }\bibfield  {title} {\enquote {\bibinfo {title} {Scaling
  theory of localization: Absence of quantum diffusion in two dimensions},}\
  }\href {https://doi.org/10.1103/PhysRevLett.42.673} {\bibfield  {journal}
  {\bibinfo  {journal} {Phys. Rev. Lett.}\ }\textbf {\bibinfo {volume} {42}},\
  \bibinfo {pages} {673} (\bibinfo {year} {1979})}\BibitemShut {NoStop}%
\bibitem [{\citenamefont {Andreani}, \citenamefont {Tassone},\ and\
  \citenamefont {Bassani}(1991)}]{Andreani1991}%
  \BibitemOpen
  \bibfield  {author} {\bibinfo {author} {\bibfnamefont {L.~C.}\ \bibnamefont
  {Andreani}}, \bibinfo {author} {\bibfnamefont {F.}~\bibnamefont {Tassone}},\
  and\ \bibinfo {author} {\bibfnamefont {F.}~\bibnamefont {Bassani}},\
  }\bibfield  {title} {\enquote {\bibinfo {title} {Radiative lifetime of free
  excitons in quantum wells},}\ }\href
  {https://doi.org/https://doi.org/10.1016/0038-1098(91)90761-J} {\bibfield
  {journal} {\bibinfo  {journal} {Solid State Communications}\ }\textbf
  {\bibinfo {volume} {77}},\ \bibinfo {pages} {641} (\bibinfo {year}
  {1991})}\BibitemShut {NoStop}%
\bibitem [{\citenamefont {Andreani}(1995)}]{Andreani-book}%
  \BibitemOpen
  \bibfield  {author} {\bibinfo {author} {\bibfnamefont {L.~C.}\ \bibnamefont
  {Andreani}},\ }\bibfield  {title} {\enquote {\bibinfo {title} {Optical
  transitions, excitons, and polaritons in bulk and low-dimensional
  semiconductor structures},}\ }in\ \href
  {https://doi.org/10.1007/978-1-4615-1963-8_3} {\emph {\bibinfo {booktitle}
  {Confined Electrons and Photons}}},\ \bibinfo {series and number} {NATO ASI
  Series},\ \bibinfo {editor} {edited by\ \bibinfo {editor} {\bibfnamefont
  {E.}~\bibnamefont {Burstein}}\ and\ \bibinfo {editor} {\bibfnamefont
  {C.}~\bibnamefont {Weisbuch}}}\ (\bibinfo  {publisher} {Springer US},\
  \bibinfo {year} {1995})\ pp.\ \bibinfo {pages} {57--112}\BibitemShut
  {NoStop}%
\bibitem [{\citenamefont {Akiyama}\ \emph {et~al.}(1994)\citenamefont
  {Akiyama}, \citenamefont {Koshiba}, \citenamefont {Someya}, \citenamefont
  {Wada}, \citenamefont {Noge}, \citenamefont {Nakamura}, \citenamefont
  {Inoshita}, \citenamefont {Shimizu},\ and\ \citenamefont
  {Sakaki}}]{Akiyama1994}%
  \BibitemOpen
  \bibfield  {author} {\bibinfo {author} {\bibfnamefont {H.}~\bibnamefont
  {Akiyama}}, \bibinfo {author} {\bibfnamefont {S.}~\bibnamefont {Koshiba}},
  \bibinfo {author} {\bibfnamefont {T.}~\bibnamefont {Someya}}, \bibinfo
  {author} {\bibfnamefont {K.}~\bibnamefont {Wada}}, \bibinfo {author}
  {\bibfnamefont {H.}~\bibnamefont {Noge}}, \bibinfo {author} {\bibfnamefont
  {Y.}~\bibnamefont {Nakamura}}, \bibinfo {author} {\bibfnamefont
  {T.}~\bibnamefont {Inoshita}}, \bibinfo {author} {\bibfnamefont
  {A.}~\bibnamefont {Shimizu}},\ and\ \bibinfo {author} {\bibfnamefont
  {H.}~\bibnamefont {Sakaki}},\ }\bibfield  {title} {\enquote {\bibinfo {title}
  {Thermalization effect on radiative decay of excitons in quantum wires},}\
  }\href {https://doi.org/10.1103/PhysRevLett.72.924} {\bibfield  {journal}
  {\bibinfo  {journal} {Phys. Rev. Lett.}\ }\textbf {\bibinfo {volume} {72}},\
  \bibinfo {pages} {924} (\bibinfo {year} {1994})}\BibitemShut {NoStop}%
\bibitem [{\citenamefont {Lomascolo}\ \emph {et~al.}(1998)\citenamefont
  {Lomascolo}, \citenamefont {Ciccarese}, \citenamefont {Cingolani},
  \citenamefont {Rinaldi},\ and\ \citenamefont {Reinhart}}]{Lomascolo1998}%
  \BibitemOpen
  \bibfield  {author} {\bibinfo {author} {\bibfnamefont {M.}~\bibnamefont
  {Lomascolo}}, \bibinfo {author} {\bibfnamefont {P.}~\bibnamefont
  {Ciccarese}}, \bibinfo {author} {\bibfnamefont {R.}~\bibnamefont
  {Cingolani}}, \bibinfo {author} {\bibfnamefont {R.}~\bibnamefont {Rinaldi}},\
  and\ \bibinfo {author} {\bibfnamefont {F.~K.}\ \bibnamefont {Reinhart}},\
  }\bibfield  {title} {\enquote {\bibinfo {title} {Free versus localized
  exciton in {GaAs} {V}-shaped quantum wires},}\ }\href
  {https://doi.org/10.1063/1.366683} {\bibfield  {journal} {\bibinfo  {journal}
  {Journal of Applied Physics}\ }\textbf {\bibinfo {volume} {83}},\ \bibinfo
  {pages} {302} (\bibinfo {year} {1998})}\BibitemShut {NoStop}%
\bibitem [{\citenamefont {Cade}\ \emph {et~al.}(2004)\citenamefont {Cade},
  \citenamefont {Roshan}, \citenamefont {Hauert}, \citenamefont {Maciel},
  \citenamefont {Ryan}, \citenamefont {Schwarz}, \citenamefont {Sch\"apers},\
  and\ \citenamefont {L\"uth}}]{Cade2004}%
  \BibitemOpen
  \bibfield  {author} {\bibinfo {author} {\bibfnamefont {N.~I.}\ \bibnamefont
  {Cade}}, \bibinfo {author} {\bibfnamefont {R.}~\bibnamefont {Roshan}},
  \bibinfo {author} {\bibfnamefont {M.}~\bibnamefont {Hauert}}, \bibinfo
  {author} {\bibfnamefont {A.~C.}\ \bibnamefont {Maciel}}, \bibinfo {author}
  {\bibfnamefont {J.~F.}\ \bibnamefont {Ryan}}, \bibinfo {author}
  {\bibfnamefont {A.}~\bibnamefont {Schwarz}}, \bibinfo {author} {\bibfnamefont
  {T.}~\bibnamefont {Sch\"apers}},\ and\ \bibinfo {author} {\bibfnamefont
  {H.}~\bibnamefont {L\"uth}},\ }\bibfield  {title} {\enquote {\bibinfo {title}
  {Carrier relaxation in {GaAs} v-groove quantum wires and the effects of
  localization},}\ }\href {https://doi.org/10.1103/PhysRevB.70.195308}
  {\bibfield  {journal} {\bibinfo  {journal} {Phys. Rev. B}\ }\textbf {\bibinfo
  {volume} {70}},\ \bibinfo {pages} {195308} (\bibinfo {year}
  {2004})}\BibitemShut {NoStop}%
\end{thebibliography}

\begin{thebibliography}{12}%
\makeatletter
\providecommand \@ifxundefined [1]{%
 \@ifx{#1\undefined}
}%
\providecommand \@ifnum [1]{%
 \ifnum #1\expandafter \@firstoftwo
 \else \expandafter \@secondoftwo
 \fi
}%
\providecommand \@ifx [1]{%
 \ifx #1\expandafter \@firstoftwo
 \else \expandafter \@secondoftwo
 \fi
}%
\providecommand \natexlab [1]{#1}%
\providecommand \enquote  [1]{``#1''}%
\providecommand \bibnamefont  [1]{#1}%
\providecommand \bibfnamefont [1]{#1}%
\providecommand \citenamefont [1]{#1}%
\providecommand \href@noop [0]{\@secondoftwo}%
\providecommand \href [0]{\begingroup \@sanitize@url \@href}%
\providecommand \@href[1]{\@@startlink{#1}\@@href}%
\providecommand \@@href[1]{\endgroup#1\@@endlink}%
\providecommand \@sanitize@url [0]{\catcode `\\12\catcode `\$12\catcode
  `\&12\catcode `\#12\catcode `\^12\catcode `\_12\catcode `\%12\relax}%
\providecommand \@@startlink[1]{}%
\providecommand \@@endlink[0]{}%
\providecommand \url  [0]{\begingroup\@sanitize@url \@url }%
\providecommand \@url [1]{\endgroup\@href {#1}{\urlprefix }}%
\providecommand \urlprefix  [0]{URL }%
\providecommand \Eprint [0]{\href }%
\providecommand \doibase [0]{https://doi.org/}%
\providecommand \selectlanguage [0]{\@gobble}%
\providecommand \bibinfo  [0]{\@secondoftwo}%
\providecommand \bibfield  [0]{\@secondoftwo}%
\providecommand \translation [1]{[#1]}%
\providecommand \BibitemOpen [0]{}%
\providecommand \bibitemStop [0]{}%
\providecommand \bibitemNoStop [0]{.\EOS\space}%
\providecommand \EOS [0]{\spacefactor3000\relax}%
\providecommand \BibitemShut  [1]{\csname bibitem#1\endcsname}%
\let\auto@bib@innerbib\@empty
\bibitem [{\citenamefont {Giannozzi}\ \emph {et~al.}(2017)\citenamefont
  {Giannozzi}, \citenamefont {Andreussi}, \citenamefont {Brumme}, \citenamefont
  {Bunau}, \citenamefont {Nardelli}, \citenamefont {Calandra}, \citenamefont
  {Car}, \citenamefont {Cavazzoni}, \citenamefont {Ceresoli}, \citenamefont
  {Cococcioni}, \citenamefont {Colonna}, \citenamefont {Carnimeo},
  \citenamefont {Corso}, \citenamefont {de~Gironcoli}, \citenamefont {Delugas},
  \citenamefont {Jr}, \citenamefont {Ferretti}, \citenamefont {Floris},
  \citenamefont {Fratesi}, \citenamefont {Fugallo}, \citenamefont {Gebauer},
  \citenamefont {Gerstmann}, \citenamefont {Giustino}, \citenamefont {Gorni},
  \citenamefont {Jia}, \citenamefont {Kawamura}, \citenamefont {Ko},
  \citenamefont {Kokalj}, \citenamefont {Küçükbenli}, \citenamefont
  {Lazzeri}, \citenamefont {Marsili}, \citenamefont {Marzari}, \citenamefont
  {Mauri}, \citenamefont {Nguyen}, \citenamefont {Nguyen}, \citenamefont {de-la
  Roza}, \citenamefont {Paulatto}, \citenamefont {Poncé}, \citenamefont
  {Rocca}, \citenamefont {Sabatini}, \citenamefont {Santra}, \citenamefont
  {Schlipf}, \citenamefont {Seitsonen}, \citenamefont {Smogunov}, \citenamefont
  {Timrov}, \citenamefont {Thonhauser}, \citenamefont {Umari}, \citenamefont
  {Vast}, \citenamefont {Wu},\ and\ \citenamefont {Baroni}}]{QE-2017}%
  \BibitemOpen
  \bibfield  {author} {\bibinfo {author} {\bibfnamefont {P.}~\bibnamefont
  {Giannozzi}}, \bibinfo {author} {\bibfnamefont {O.}~\bibnamefont
  {Andreussi}}, \bibinfo {author} {\bibfnamefont {T.}~\bibnamefont {Brumme}},
  \bibinfo {author} {\bibfnamefont {O.}~\bibnamefont {Bunau}}, \bibinfo
  {author} {\bibfnamefont {M.~B.}\ \bibnamefont {Nardelli}}, \bibinfo {author}
  {\bibfnamefont {M.}~\bibnamefont {Calandra}}, \bibinfo {author}
  {\bibfnamefont {R.}~\bibnamefont {Car}}, \bibinfo {author} {\bibfnamefont
  {C.}~\bibnamefont {Cavazzoni}}, \bibinfo {author} {\bibfnamefont
  {D.}~\bibnamefont {Ceresoli}}, \bibinfo {author} {\bibfnamefont
  {M.}~\bibnamefont {Cococcioni}}, \bibinfo {author} {\bibfnamefont
  {N.}~\bibnamefont {Colonna}}, \bibinfo {author} {\bibfnamefont
  {I.}~\bibnamefont {Carnimeo}}, \bibinfo {author} {\bibfnamefont {A.~D.}\
  \bibnamefont {Corso}}, \bibinfo {author} {\bibfnamefont {S.}~\bibnamefont
  {de~Gironcoli}}, \bibinfo {author} {\bibfnamefont {P.}~\bibnamefont
  {Delugas}}, \bibinfo {author} {\bibfnamefont {R.~A.~D.}\ \bibnamefont {Jr}},
  \bibinfo {author} {\bibfnamefont {A.}~\bibnamefont {Ferretti}}, \bibinfo
  {author} {\bibfnamefont {A.}~\bibnamefont {Floris}}, \bibinfo {author}
  {\bibfnamefont {G.}~\bibnamefont {Fratesi}}, \bibinfo {author} {\bibfnamefont
  {G.}~\bibnamefont {Fugallo}}, \bibinfo {author} {\bibfnamefont
  {R.}~\bibnamefont {Gebauer}}, \bibinfo {author} {\bibfnamefont
  {U.}~\bibnamefont {Gerstmann}}, \bibinfo {author} {\bibfnamefont
  {F.}~\bibnamefont {Giustino}}, \bibinfo {author} {\bibfnamefont
  {T.}~\bibnamefont {Gorni}}, \bibinfo {author} {\bibfnamefont
  {J.}~\bibnamefont {Jia}}, \bibinfo {author} {\bibfnamefont {M.}~\bibnamefont
  {Kawamura}}, \bibinfo {author} {\bibfnamefont {H.-Y.}\ \bibnamefont {Ko}},
  \bibinfo {author} {\bibfnamefont {A.}~\bibnamefont {Kokalj}}, \bibinfo
  {author} {\bibfnamefont {E.}~\bibnamefont {Küçükbenli}}, \bibinfo {author}
  {\bibfnamefont {M.}~\bibnamefont {Lazzeri}}, \bibinfo {author} {\bibfnamefont
  {M.}~\bibnamefont {Marsili}}, \bibinfo {author} {\bibfnamefont
  {N.}~\bibnamefont {Marzari}}, \bibinfo {author} {\bibfnamefont
  {F.}~\bibnamefont {Mauri}}, \bibinfo {author} {\bibfnamefont {N.~L.}\
  \bibnamefont {Nguyen}}, \bibinfo {author} {\bibfnamefont {H.-V.}\
  \bibnamefont {Nguyen}}, \bibinfo {author} {\bibfnamefont {A.~O.}\
  \bibnamefont {de-la Roza}}, \bibinfo {author} {\bibfnamefont
  {L.}~\bibnamefont {Paulatto}}, \bibinfo {author} {\bibfnamefont
  {S.}~\bibnamefont {Poncé}}, \bibinfo {author} {\bibfnamefont
  {D.}~\bibnamefont {Rocca}}, \bibinfo {author} {\bibfnamefont
  {R.}~\bibnamefont {Sabatini}}, \bibinfo {author} {\bibfnamefont
  {B.}~\bibnamefont {Santra}}, \bibinfo {author} {\bibfnamefont
  {M.}~\bibnamefont {Schlipf}}, \bibinfo {author} {\bibfnamefont {A.~P.}\
  \bibnamefont {Seitsonen}}, \bibinfo {author} {\bibfnamefont {A.}~\bibnamefont
  {Smogunov}}, \bibinfo {author} {\bibfnamefont {I.}~\bibnamefont {Timrov}},
  \bibinfo {author} {\bibfnamefont {T.}~\bibnamefont {Thonhauser}}, \bibinfo
  {author} {\bibfnamefont {P.}~\bibnamefont {Umari}}, \bibinfo {author}
  {\bibfnamefont {N.}~\bibnamefont {Vast}}, \bibinfo {author} {\bibfnamefont
  {X.}~\bibnamefont {Wu}},\ and\ \bibinfo {author} {\bibfnamefont
  {S.}~\bibnamefont {Baroni}},\ }\bibfield  {title} {\enquote {\bibinfo {title}
  {Advanced capabilities for materials modelling with {QUANTUM ESPRESSO}},}\
  }\href {http://stacks.iop.org/0953-8984/29/i=46/a=465901} {\bibfield
  {journal} {\bibinfo  {journal} {Journal of Physics: Condensed Matter}\
  }\textbf {\bibinfo {volume} {29}},\ \bibinfo {pages} {465901} (\bibinfo
  {year} {2017})}\BibitemShut {NoStop}%
\bibitem [{\citenamefont {Hamann}(2013)}]{hamann2013optimized}%
  \BibitemOpen
  \bibfield  {author} {\bibinfo {author} {\bibfnamefont {D.}~\bibnamefont
  {Hamann}},\ }\bibfield  {title} {\enquote {\bibinfo {title} {Optimized
  norm-conserving {V}anderbilt pseudopotentials},}\ }\href@noop {} {\bibfield
  {journal} {\bibinfo  {journal} {Physical Review B—Condensed Matter and
  Materials Physics}\ }\textbf {\bibinfo {volume} {88}},\ \bibinfo {pages}
  {085117} (\bibinfo {year} {2013})}\BibitemShut {NoStop}%
\bibitem [{\citenamefont {Telford}\ \emph {et~al.}(2020)\citenamefont
  {Telford}, \citenamefont {Dismukes}, \citenamefont {Lee}, \citenamefont
  {Cheng}, \citenamefont {Wieteska}, \citenamefont {Bartholomew}, \citenamefont
  {Chen}, \citenamefont {Xu}, \citenamefont {Pasupathy}, \citenamefont {Zhu}
  \emph {et~al.}}]{telford2020layered}%
  \BibitemOpen
  \bibfield  {author} {\bibinfo {author} {\bibfnamefont {E.~J.}\ \bibnamefont
  {Telford}}, \bibinfo {author} {\bibfnamefont {A.~H.}\ \bibnamefont
  {Dismukes}}, \bibinfo {author} {\bibfnamefont {K.}~\bibnamefont {Lee}},
  \bibinfo {author} {\bibfnamefont {M.}~\bibnamefont {Cheng}}, \bibinfo
  {author} {\bibfnamefont {A.}~\bibnamefont {Wieteska}}, \bibinfo {author}
  {\bibfnamefont {A.~K.}\ \bibnamefont {Bartholomew}}, \bibinfo {author}
  {\bibfnamefont {Y.-S.}\ \bibnamefont {Chen}}, \bibinfo {author}
  {\bibfnamefont {X.}~\bibnamefont {Xu}}, \bibinfo {author} {\bibfnamefont
  {A.~N.}\ \bibnamefont {Pasupathy}}, \bibinfo {author} {\bibfnamefont
  {X.}~\bibnamefont {Zhu}}, \emph {et~al.},\ }\bibfield  {title} {\enquote
  {\bibinfo {title} {Layered antiferromagnetism induces large negative
  magnetoresistance in the van der waals semiconductor crsbr},}\ }\href@noop {}
  {\bibfield  {journal} {\bibinfo  {journal} {Advanced Materials}\ }\textbf
  {\bibinfo {volume} {32}},\ \bibinfo {pages} {2003240} (\bibinfo {year}
  {2020})}\BibitemShut {NoStop}%
\bibitem [{\citenamefont {Wang}, \citenamefont {Qi},\ and\ \citenamefont
  {Qian}(2020)}]{wang2020electrically}%
  \BibitemOpen
  \bibfield  {author} {\bibinfo {author} {\bibfnamefont {H.}~\bibnamefont
  {Wang}}, \bibinfo {author} {\bibfnamefont {J.}~\bibnamefont {Qi}},\ and\
  \bibinfo {author} {\bibfnamefont {X.}~\bibnamefont {Qian}},\ }\bibfield
  {title} {\enquote {\bibinfo {title} {Electrically tunable high {C}urie
  temperature two-dimensional ferromagnetism in van der {W}aals layered
  crystals},}\ }\href@noop {} {\bibfield  {journal} {\bibinfo  {journal}
  {Applied Physics Letters}\ }\textbf {\bibinfo {volume} {117}} (\bibinfo
  {year} {2020})}\BibitemShut {NoStop}%
\bibitem [{\citenamefont {Lin}\ \emph {et~al.}(2024)\citenamefont {Lin},
  \citenamefont {Li}, \citenamefont {Ghorbani-Asl}, \citenamefont {Sofer},
  \citenamefont {Winnerl}, \citenamefont {Erbe}, \citenamefont
  {Krasheninnikov}, \citenamefont {Helm}, \citenamefont {Zhou}, \citenamefont
  {Dan} \emph {et~al.}}]{Lin-2024b}%
  \BibitemOpen
  \bibfield  {author} {\bibinfo {author} {\bibfnamefont {K.}~\bibnamefont
  {Lin}}, \bibinfo {author} {\bibfnamefont {Y.}~\bibnamefont {Li}}, \bibinfo
  {author} {\bibfnamefont {M.}~\bibnamefont {Ghorbani-Asl}}, \bibinfo {author}
  {\bibfnamefont {Z.}~\bibnamefont {Sofer}}, \bibinfo {author} {\bibfnamefont
  {S.}~\bibnamefont {Winnerl}}, \bibinfo {author} {\bibfnamefont
  {A.}~\bibnamefont {Erbe}}, \bibinfo {author} {\bibfnamefont {A.~V.}\
  \bibnamefont {Krasheninnikov}}, \bibinfo {author} {\bibfnamefont
  {M.}~\bibnamefont {Helm}}, \bibinfo {author} {\bibfnamefont {S.}~\bibnamefont
  {Zhou}}, \bibinfo {author} {\bibfnamefont {Y.}~\bibnamefont {Dan}}, \emph
  {et~al.},\ }\bibfield  {title} {\enquote {\bibinfo {title} {Probing the band
  splitting near the {$\Gamma$} point in the van der waals magnetic
  semiconductor {CrSBr}},}\ }\href
  {https://doi.org/10.1021/acs.jpclett.4c00968} {\bibfield  {journal} {\bibinfo
   {journal} {The Journal of Physical Chemistry Letters}\ }\textbf {\bibinfo
  {volume} {15}},\ \bibinfo {pages} {6010} (\bibinfo {year}
  {2024})}\BibitemShut {NoStop}%
\bibitem [{\citenamefont {Sangalli}\ \emph {et~al.}(2019)\citenamefont
  {Sangalli}, \citenamefont {Ferretti}, \citenamefont {Miranda}, \citenamefont
  {Attaccalite}, \citenamefont {Marri}, \citenamefont {Cannuccia},
  \citenamefont {Melo}, \citenamefont {Marsili}, \citenamefont {Paleari},
  \citenamefont {Marrazzo} \emph {et~al.}}]{sangalli2019many}%
  \BibitemOpen
  \bibfield  {author} {\bibinfo {author} {\bibfnamefont {D.}~\bibnamefont
  {Sangalli}}, \bibinfo {author} {\bibfnamefont {A.}~\bibnamefont {Ferretti}},
  \bibinfo {author} {\bibfnamefont {H.}~\bibnamefont {Miranda}}, \bibinfo
  {author} {\bibfnamefont {C.}~\bibnamefont {Attaccalite}}, \bibinfo {author}
  {\bibfnamefont {I.}~\bibnamefont {Marri}}, \bibinfo {author} {\bibfnamefont
  {E.}~\bibnamefont {Cannuccia}}, \bibinfo {author} {\bibfnamefont
  {P.}~\bibnamefont {Melo}}, \bibinfo {author} {\bibfnamefont {M.}~\bibnamefont
  {Marsili}}, \bibinfo {author} {\bibfnamefont {F.}~\bibnamefont {Paleari}},
  \bibinfo {author} {\bibfnamefont {A.}~\bibnamefont {Marrazzo}}, \emph
  {et~al.},\ }\bibfield  {title} {\enquote {\bibinfo {title} {Many-body
  perturbation theory calculations using the yambo code},}\ }\href@noop {}
  {\bibfield  {journal} {\bibinfo  {journal} {Journal of physics: Condensed
  matter}\ }\textbf {\bibinfo {volume} {31}},\ \bibinfo {pages} {325902}
  (\bibinfo {year} {2019})}\BibitemShut {NoStop}%
\bibitem [{\citenamefont {Klein}\ \emph {et~al.}(2023)\citenamefont {Klein},
  \citenamefont {Pingault}, \citenamefont {Florian}, \citenamefont
  {Heißenb{\"u}ttel}, \citenamefont {Steinhoff}, \citenamefont {Song},
  \citenamefont {Torres}, \citenamefont {Dirnberger}, \citenamefont {Curtis},
  \citenamefont {Weile}, \citenamefont {Penn}, \citenamefont {Deilmann},
  \citenamefont {Dana}, \citenamefont {Bushati}, \citenamefont {Quan},
  \citenamefont {Luxa}, \citenamefont {Sofer}, \citenamefont {Alù},
  \citenamefont {Menon}, \citenamefont {Wurstbauer}, \citenamefont {Rohlfing},
  \citenamefont {Narang}, \citenamefont {Lončar},\ and\ \citenamefont
  {Ross}}]{Klein-2023}%
  \BibitemOpen
  \bibfield  {author} {\bibinfo {author} {\bibfnamefont {J.}~\bibnamefont
  {Klein}}, \bibinfo {author} {\bibfnamefont {B.}~\bibnamefont {Pingault}},
  \bibinfo {author} {\bibfnamefont {M.}~\bibnamefont {Florian}}, \bibinfo
  {author} {\bibfnamefont {M.-C.}\ \bibnamefont {Heißenb{\"u}ttel}}, \bibinfo
  {author} {\bibfnamefont {A.}~\bibnamefont {Steinhoff}}, \bibinfo {author}
  {\bibfnamefont {Z.}~\bibnamefont {Song}}, \bibinfo {author} {\bibfnamefont
  {K.}~\bibnamefont {Torres}}, \bibinfo {author} {\bibfnamefont
  {F.}~\bibnamefont {Dirnberger}}, \bibinfo {author} {\bibfnamefont {J.~B.}\
  \bibnamefont {Curtis}}, \bibinfo {author} {\bibfnamefont {M.}~\bibnamefont
  {Weile}}, \bibinfo {author} {\bibfnamefont {A.}~\bibnamefont {Penn}},
  \bibinfo {author} {\bibfnamefont {T.}~\bibnamefont {Deilmann}}, \bibinfo
  {author} {\bibfnamefont {R.}~\bibnamefont {Dana}}, \bibinfo {author}
  {\bibfnamefont {R.}~\bibnamefont {Bushati}}, \bibinfo {author} {\bibfnamefont
  {J.}~\bibnamefont {Quan}}, \bibinfo {author} {\bibfnamefont {J.}~\bibnamefont
  {Luxa}}, \bibinfo {author} {\bibfnamefont {Z.}~\bibnamefont {Sofer}},
  \bibinfo {author} {\bibfnamefont {A.}~\bibnamefont {Alù}}, \bibinfo {author}
  {\bibfnamefont {V.~M.}\ \bibnamefont {Menon}}, \bibinfo {author}
  {\bibfnamefont {U.}~\bibnamefont {Wurstbauer}}, \bibinfo {author}
  {\bibfnamefont {M.}~\bibnamefont {Rohlfing}}, \bibinfo {author}
  {\bibfnamefont {P.}~\bibnamefont {Narang}}, \bibinfo {author} {\bibfnamefont
  {M.}~\bibnamefont {Lončar}},\ and\ \bibinfo {author} {\bibfnamefont {F.~M.}\
  \bibnamefont {Ross}},\ }\bibfield  {title} {\enquote {\bibinfo {title} {The
  bulk van der {W}aals layered magnet {CrSBr} is a quasi-{1D} material},}\
  }\href {https://doi.org/10.1021/acsnano.2c07316} {\bibfield  {journal}
  {\bibinfo  {journal} {ACS Nano}\ }\textbf {\bibinfo {volume} {17}},\ \bibinfo
  {pages} {5316} (\bibinfo {year} {2023})}\BibitemShut {NoStop}%
\bibitem [{\citenamefont {Hei{\ss}enb{\"u}ttel}\ \emph
  {et~al.}(2025)\citenamefont {Hei{\ss}enb{\"u}ttel}, \citenamefont {Piel},
  \citenamefont {Klein}, \citenamefont {Deilmann}, \citenamefont {Wurstbauer},\
  and\ \citenamefont {Rohlfing}}]{heissenbuttel2025quadratic}%
  \BibitemOpen
  \bibfield  {author} {\bibinfo {author} {\bibfnamefont {M.-C.}\ \bibnamefont
  {Hei{\ss}enb{\"u}ttel}}, \bibinfo {author} {\bibfnamefont {P.-M.}\
  \bibnamefont {Piel}}, \bibinfo {author} {\bibfnamefont {J.}~\bibnamefont
  {Klein}}, \bibinfo {author} {\bibfnamefont {T.}~\bibnamefont {Deilmann}},
  \bibinfo {author} {\bibfnamefont {U.}~\bibnamefont {Wurstbauer}},\ and\
  \bibinfo {author} {\bibfnamefont {M.}~\bibnamefont {Rohlfing}},\ }\bibfield
  {title} {\enquote {\bibinfo {title} {Quadratic optical response to a magnetic
  field in the layered magnet {CrSBr}},}\ }\href@noop {} {\bibfield  {journal}
  {\bibinfo  {journal} {Physical Review B}\ }\textbf {\bibinfo {volume}
  {111}},\ \bibinfo {pages} {075107} (\bibinfo {year} {2025})}\BibitemShut
  {NoStop}%
\bibitem [{\citenamefont {Bianchi}\ \emph {et~al.}(2023)\citenamefont
  {Bianchi}, \citenamefont {Acharya}, \citenamefont {Dirnberger}, \citenamefont
  {Klein}, \citenamefont {Pashov}, \citenamefont {Mosina}, \citenamefont
  {Sofer}, \citenamefont {Rudenko}, \citenamefont {Katsnelson}, \citenamefont
  {Van~Schilfgaarde} \emph {et~al.}}]{bianchi2023paramagnetic}%
  \BibitemOpen
  \bibfield  {author} {\bibinfo {author} {\bibfnamefont {M.}~\bibnamefont
  {Bianchi}}, \bibinfo {author} {\bibfnamefont {S.}~\bibnamefont {Acharya}},
  \bibinfo {author} {\bibfnamefont {F.}~\bibnamefont {Dirnberger}}, \bibinfo
  {author} {\bibfnamefont {J.}~\bibnamefont {Klein}}, \bibinfo {author}
  {\bibfnamefont {D.}~\bibnamefont {Pashov}}, \bibinfo {author} {\bibfnamefont
  {K.}~\bibnamefont {Mosina}}, \bibinfo {author} {\bibfnamefont
  {Z.}~\bibnamefont {Sofer}}, \bibinfo {author} {\bibfnamefont {A.~N.}\
  \bibnamefont {Rudenko}}, \bibinfo {author} {\bibfnamefont {M.~I.}\
  \bibnamefont {Katsnelson}}, \bibinfo {author} {\bibfnamefont
  {M.}~\bibnamefont {Van~Schilfgaarde}}, \emph {et~al.},\ }\bibfield  {title}
  {\enquote {\bibinfo {title} {Paramagnetic electronic structure of {CrSBr}:
  Comparison between ab initio {$GW$} theory and angle-resolved photoemission
  spectroscopy},}\ }\href@noop {} {\bibfield  {journal} {\bibinfo  {journal}
  {Physical Review B}\ }\textbf {\bibinfo {volume} {107}},\ \bibinfo {pages}
  {235107} (\bibinfo {year} {2023})}\BibitemShut {NoStop}%
\bibitem [{\citenamefont {Hernandez}, \citenamefont {Roman},\ and\
  \citenamefont {Vidal}(2005)}]{hernandez2005slepc}%
  \BibitemOpen
  \bibfield  {author} {\bibinfo {author} {\bibfnamefont {V.}~\bibnamefont
  {Hernandez}}, \bibinfo {author} {\bibfnamefont {J.~E.}\ \bibnamefont
  {Roman}},\ and\ \bibinfo {author} {\bibfnamefont {V.}~\bibnamefont {Vidal}},\
  }\bibfield  {title} {\enquote {\bibinfo {title} {{SLEPc}: A scalable and
  flexible toolkit for the solution of eigenvalue problems},}\ }\href@noop {}
  {\bibfield  {journal} {\bibinfo  {journal} {ACM Transactions on Mathematical
  Software (TOMS)}\ }\textbf {\bibinfo {volume} {31}},\ \bibinfo {pages} {351}
  (\bibinfo {year} {2005})}\BibitemShut {NoStop}%
\bibitem [{\citenamefont {Fuchs}\ \emph {et~al.}(2008)\citenamefont {Fuchs},
  \citenamefont {R{\"o}dl}, \citenamefont {Schleife},\ and\ \citenamefont
  {Bechstedt}}]{fuchs2008efficient}%
  \BibitemOpen
  \bibfield  {author} {\bibinfo {author} {\bibfnamefont {F.}~\bibnamefont
  {Fuchs}}, \bibinfo {author} {\bibfnamefont {C.}~\bibnamefont {R{\"o}dl}},
  \bibinfo {author} {\bibfnamefont {A.}~\bibnamefont {Schleife}},\ and\
  \bibinfo {author} {\bibfnamefont {F.}~\bibnamefont {Bechstedt}},\ }\bibfield
  {title} {\enquote {\bibinfo {title} {Efficient {O(N2)} approach to solve the
  {B}ethe-{S}alpeter equation for excitonic bound states},}\ }\href@noop {}
  {\bibfield  {journal} {\bibinfo  {journal} {Physical Review B—Condensed
  Matter and Materials Physics}\ }\textbf {\bibinfo {volume} {78}},\ \bibinfo
  {pages} {085103} (\bibinfo {year} {2008})}\BibitemShut {NoStop}%
\bibitem [{\citenamefont {Semina}\ \emph {et~al.}(2025)\citenamefont {Semina},
  \citenamefont {Tabataba-Vakili}, \citenamefont {Rupp}, \citenamefont
  {Baimuratov}, \citenamefont {H\"ogele},\ and\ \citenamefont
  {Glazov}}]{Semina2025}%
  \BibitemOpen
  \bibfield  {author} {\bibinfo {author} {\bibfnamefont {M.~A.}\ \bibnamefont
  {Semina}}, \bibinfo {author} {\bibfnamefont {F.}~\bibnamefont
  {Tabataba-Vakili}}, \bibinfo {author} {\bibfnamefont {A.}~\bibnamefont
  {Rupp}}, \bibinfo {author} {\bibfnamefont {A.~S.}\ \bibnamefont
  {Baimuratov}}, \bibinfo {author} {\bibfnamefont {A.}~\bibnamefont
  {H\"ogele}},\ and\ \bibinfo {author} {\bibfnamefont {M.~M.}\ \bibnamefont
  {Glazov}},\ }\bibfield  {title} {\enquote {\bibinfo {title} {Excitons and
  trions in {CrSBr} bilayers},}\ }\href
  {https://doi.org/10.1103/PhysRevB.111.205301} {\bibfield  {journal} {\bibinfo
   {journal} {Phys. Rev. B}\ }\textbf {\bibinfo {volume} {111}},\ \bibinfo
  {pages} {205301} (\bibinfo {year} {2025})}\BibitemShut {NoStop}%
\end{thebibliography}
\end{document}